\documentclass[12pt,preprint]{aastex}

\usepackage{natbib}
\usepackage{graphicx}
\newcommand{\eqb}{\begin{equation}}
\newcommand{\eqe}{\end{equation}}
\begin{document}

\title{Asymptotic structure of Poynting dominated jets}

\author{Yuri Lyubarsky}
\affil{Physics Department, Ben-Gurion University, P.O.B. 653, Beer-Sheva 84105, Israel}

\begin{abstract}
In relativistic, Poynting dominated outflows, acceleration and collimation are intimately connected.
An important point is that
the Lorentz force is nearly compensated by the electric force
therefore the acceleration zone spans a large range of scales.
We derived the asymptotic equations describing relativistic,
axisymmetric MHD flows far beyond the light cylinder. These
equations do not contain either intrinsic small scales (like
the light cylinder radius) or terms that nearly cancel each
other (like the electric and magnetic forces) therefore they
could be easily solved numerically. They also suit well for
qualitative analysis of the flow and in many cases, they could
even be solved analytically or semi-analytically. We show that
there are generally two collimation regimes. In the first regime,
the residual of the hoop stress and the electric force is
counterbalanced by the pressure of the poloidal magnetic field
so that at any distance from the source, the structure of the
flow is the same as the structure of an appropriate cylindrical
equilibrium configuration. In the second regime, the pressure of
the poloidal magnetic field is negligible small so that the
flow could be conceived as composed from coaxial magnetic loops.
In the two collimation regimes, the flow is accelerated in different ways. We study in detail the
structure of jets confined by the external pressure with a
power law profile. In particular, we obtained simple scalings
for the extent of the acceleration zone, for the terminal
Lorentz factor and for the collimation angle.
\end{abstract}

\keywords{(magnetohydrodynamics:) MHD -- relativity -- galaxies:jets -- gamma rays: bursts}

\maketitle

\section{Introduction}
Highly collimated, relativistic jets are observed in active
galactic nuclei (AGNs), microquasars and gamma-ray bursts
(GRBs). According to the most popular model, these outflows are
powered hydromagnetically. By analogy with pulsars, it is
assumed that the magnetosphere of a rapidly rotating accretion
disk \citep{lovelace76,blandford76} and the black hole itself
\citep{blandford_znajek77} opens into a relativistic wind that
transfers the energy away in the form of the Poynting flux. A
long debated question is how and where the electro-magnetic
energy is transferred to the plasma. The Poynting flux could be
transferred to the kinetic energy of the flow by gradual
acceleration however, the acceleration strongly depends on the
geometry of the flow \citep{chiueh_li_begelman91,begelman_li94,vlahakis04}
so that acceleration and collimation are intimately connected.

General theorems affirm
\citep{heyvaerts_norman89,heyvaerts_norman03,chiueh_li_begelman91,bogovalov95}
that at the infinity, the flow should collimate to the rotational
axis, a good fraction of the electromagnetic energy being
converted into the kinetic energy.  
However, it has been found that without an external
confinement, the characteristic collimation/acceleration scale
is exponentially large
\citep{eichler93,begelman_li94,tomimatsu94,beskin_etal98,bogovalov98,chieh_etal98,bogovalov_tsinganos99}.
That is why in pulsar winds, the Poynting flux is converted
into the plasma energy predominantly via dissipation processes
(see, e.g., review by \citet{kirk_etal07}). On the other hand,
relativistic jets are observed in the sources where 
interaction of the outflows with the external medium could not
be neglected. 
In accreting systems, the relativistic outflows from the black
hole and the internal part of the accretion disc could be
confined by the (generally magnetized) wind from the outer
parts of the disk. A widely accepted model of long-duration
GRBs assumes that a relativistic jet from the collapsing core
pushes its way through the stellar envelope. In all these cases
the external pressure could be responsible for collimation of
Poynting dominated outflows. Moreover, the flow is efficiently
accelerated in the collimated outflows so that a significant
fraction of the Poynting flux could be eventually converted
into the plasma kinetic energy. Note that non-magnetized jets
could also be efficiently focused by an ambient medium
\citep{eichler82,peter_eichler95,levinson_eichler00,bromberg_levinson07}.
An advantage of magnetically driven outflows is a relatively
low mass load, which naturally leads to highly  relativistic
velocities.

An explicit solution for the relativistic magnetized wind from
the accretion disk was found in the force-free approximation by
\citet{blandford76}. In this solution, the magnetic surfaces
are nested paraboloids. \citet{beskin_nokhrina06}  generalized
this solution to include the inertia forces and showed that the
magnetic surfaces are only slightly modified and that the flow
is accelerated until the equipartition
level. A few self-similar solutions to the relativistic
magnetohydrodynamic (MHD) equations were found
\citep{li_chiueh_begelman92,contopoulos95,vlahakis_konigl03a,vlahakis_konigl03b,narayan_etal07},
which resemble outflows from a disk. These solutions also
demonstrated that collimation and acceleration could occur at a
reasonable, even though large, scale. 
Numerical simulations support these findings \citep{komissarov_etal07,komissarov_etal08,tchekhovskoy08}.

A crucial assumption in these models is a non-zero magnetic
flux threading the disk and the black hole. 
The total flux should in fact be infinite (going to infinity with the outer disk
radius) 
because it is the pressure
of the poloidal field, not the hoop stress, that collimates the outflow
\citep{spruit_etal97}. Such a field could not be generated in
the disk; it should be dragged inward by the accreting material
\citep{bisnovaty_ruzmaikin76,bisnovaty_lovelace07,rothstein_lovelace08}.
Magnetized outflows with the zero net magnetic flux, the so
called magnetic towers, were proposed by \citet{lynden-bell96}
and then studied both analytically
\citep{lovelace_romanova03,uzdensky_macfadyen06,lynden-bell06,lynden-bell07,gourgouliatos_lynden-bell08}
and numerically
\citep{lovelace_etal02,kato_etal04,nakamura_etal06,nakamura_etal07}.
Since there is no large scale magnetic field in this model, the jet is collimated by the pressure of the ambient medium so that an extended outflow
surrounding the jet is anyway needed.

It is well known that in relativistic MHD outflows, the
acceleration zone spans a large range of scales.
This is because the electric force, which is negligibly small
in the non-relativistic case, becomes comparable with the
Lorentz force and when the flow velocity approaches the speed
of light, these two forces nearly cancel each other so that
both acceleration and collimation proceed very slowly. Within
the light cylinder\footnote{In differentially rotating
magnetospheres, the surface $\Omega r=1$ is not a cylinder but
we retain the standard term, which has come from the pulsar
theory.},
the magnetosphere corotates with the
central source so that the plasma, which slides along the rotating
field lines, could acquire only moderate relativistic
velocities.
Beyond the light cylinder, the flow is accelerated at least until
the velocity exceeds the fast magnetosonic velocity.
The fast magnetosonic point is already very far from the light
cylinder but in this point, the plasma energy is still well
below the Poynting flux (e.g., \citet{camenzind86}). The
complete transformation of the electro-magnetic to the kinetic
energy could occur only at the scale much larger than even the
distance to the fast
magnetosonic surface. This means that a few different spatial scales
are present in the problem, which poses a strong challenge to numerical simulations. 
On the other hand, multi-scale systems
are suitable for asymptotic analysis. In
the spirit of the method of matched asymptotic expansions, one
can solve the equations in two overlapping domains, namely, in
the near zone, $\Omega r\sim 1$, where the force-free
approximation could be used, and in the far zone, where one can
considerably simplify the equations in the limit $(\Omega
r)^{-1}\ll 1$. Both solutions should be matched in the
intermediate region where the flow is still force-free but the
condition $\Omega r\gg 1$ is already
fulfilled. 

In this paper, we study properties of relativistic jets at the
distances much larger than the light cylinder radius. First we
obtain the asymptotic equations describing the flow in the
limit $\Omega r\gg 1$. Far enough from the source, these
equations are valid till the axis of the flow so that these
equations in fact describe the whole flow in the far zone. We
apply the obtained equations to jets confined by an ambient
medium. We show that there are two different regimes of the flow
collimation and acceleration. In the first regime, the structure
of the flow at any distance from the source is the same as in
an appropriate cylindrical jet, i.e., the residual between the
magnetic hoop stress and the electric force is compensated by
the pressure of the poloidal field. We will refer to this regime
as to equilibrium collimation in the sense that the flow
remains in the cylindrical equilibrium. In the second regime, one
can neglect the pressure of the poloidal field so that the dynamics
of the flow is the same as in the case of purely toroidal field;
this regime will be called non-equilibrium. In different collimation regimes, the acceleration
regimes are also different.

We show that while
the flow is Poynting dominated, the structure of the jet is
governed by a simple ordinary differential equation, which
could be easily solved for any distribution of the external
pressure. The general theory will be applied to jets with a
constant angular velocity propagating in a medium with the
pressure decreasing as a power law. We also study the structure
of the moderately magnetized core of the jet; such a core is
presented near the axis of even Poynting dominated flows
because the Poynting flux vanishes at the axis. As the jet
propagates, the flow is accelerated and the inner parts of the jet reach equipartition
between the kinetic and electromagnetic energy so that the the moderately magnetized
core expands within the jet. Depending on the
profile of the confining pressure, the core could either
occupy only internal part of the jet so that the main body of the flow remains
Poynting dominated or expand till the periphery of the flow such that the
whole jet ceases to be Poynting dominated.

The paper is organized  as follows. In the next section, we
shortly outline derivation of the basic equations describing
relativistic, axisymmetric MHD flows. In Sect. 3, we shortly
discuss the boundary conditions and integrals of motions. In
Sect. 4, we find asymptotic equations for the flow in the far
zone. In Sect. 5, we use the derived equations to develop a
technique for finding the structure of collimated, Poynting
dominated jets. In Sect. 6, we apply this technique to jets
with a constant angular velocity propagating in a medium with
the pressure decreasing as a power law. The terminal Lorentz
factor of the flow as well the terminal collimation angle, are
estimated in Sect. 7. In Sect. 8, we study the structure of the
moderately magnetized core of the jet. The obtained results are
summarized in Sect. 9.



\section{Basic equations}

For the sake of consistency and in order to introduce
notations, let us shortly review the basic theory of relativistic,
magnetized winds \citep{okamoto78,lovelace_etal86,li_chiueh_begelman92}.
Let the plasma be cold, which is a good approximation in the far zone where the
flow is already expanded. Then the steady state equation of motion is written as
 \eqb
\rho\gamma(\mathbf{v\cdot\nabla})\gamma\mathbf{v}=
\frac 1{4\pi}\left[( \nabla\cdot\mathbf{E})\mathbf{E}+\mathbf{(\nabla\times B)\times B}\right];
 \eqe
where $\rho$ is the plasma proper density, $\gamma$ the Lorentz factor, $\mathbf{v}$ the plasma velocity;
the speed of light is taken to be unity. Here the second pair of Maxwell's equation is
already used. The equation of motion should be
supplemented by the first pair of Maxwell's equations,
 \eqb
\nabla\cdot\mathbf{B}=0;\quad \mathbf{\nabla\times E}=0;
 \label{maxwell}\eqe
by the continuity equation,
 \eqb
\nabla\cdot(\rho\gamma\mathbf{v})=0;
 \label{cont}\eqe
and by the condition of flux freezing,
 \eqb
 \mathbf{E}+\mathbf{v\times B}=0.
 \label{freezing}\eqe

In axisymmetric configurations, the magnetic field is conveniently decomposed into the poloidal and
toroidal components, $\mathbf{B}=\mathbf{B}_p+B_{\phi}\mathbf{\widehat{\phi}}$, the poloidal field
being expressed via the flux function
 \eqb
 \mathbf{B}_p=\frac 1r\nabla\Psi\times\mathbf{\widehat{\phi}}.
 \label{Bfield}\eqe
Here cylindrical $(r,\phi,z)$ coordinates are used; hat denotes unite vectors. The
condition of flux freezing implies that the flux surfaces are equipotentials, which yields
 \eqb
\mathbf{E}=-\Omega(\Psi)\nabla\Psi;
 \label{Efield}\eqe
where $\Omega(\Psi)$ is the angular velocity of the field line. This gives a
 useful relation
 \eqb
E=r\Omega B_p.
 \label{EBp}\eqe
The plasma streams along the flux surfaces so that the flow velocity may also  be decomposed into the
poloidal and toroidal components, $\mathbf{v}=v_p\mathbf{\widehat{l}}+v_{\phi}\widehat{\phi}$, where
$\mathbf{\widehat{l}}$ is the unit vector along the magnetic surface,
 \eqb
\mathbf{\widehat{l}}=\mathbf{\widehat{n}\times\widehat{\phi}};\quad
\mathbf{\widehat{n}}=\nabla\Psi/\vert\nabla\Psi\vert.
 \eqe
The condition of flux freezing yields a relation between the components of the velocity and magnetic field:
 \eqb
B_pv_{\phi}-B_{\phi}v_p=r\Omega(\Psi) B_p;
 \label{Omega}\eqe
which implies that the plasma slides along the rotating magnetic field lines.
The continuity equation (\ref{cont}) could be integrated, with the aid of Eq. (\ref{maxwell}), into the
conservation law
 \eqb
4\pi\rho v_p\gamma=\eta(\Psi)B_p;
 \label{continuity}\eqe
where the function $\eta$ describes the distribution of the mass flux at the inlet of the flow.

The three remaining equations are obtained by projecting the equation of motion
onto directions $\mathbf{\widehat{l}}$, $\mathbf{\widehat{\phi}}$ and $\mathbf{\widehat{n}}$. The first
two may be manipulated into the integrals of motion
 \eqb
 \gamma-\frac{r\Omega B_{\phi}}{\eta}=\mu(\Psi);
 \label{energy}\eqe
 \eqb
 \gamma rv_{\phi}-\frac{rB_{\phi}}{\eta}=l(\Psi);
 \label{momentum}\eqe
representing conservation of the energy and of the angular
momentum, correspondingly. Note that the widely used parameter
$\sigma$, defined as the ratio of the Poynting to the matter
energy flux, is presented via the basic quantities as
 \eqb
\sigma=\frac{\mu-\gamma}{\gamma}.
 \eqe
The projection of the equation of motion onto the normal to the
flux surface, $\mathbf{\widehat{n}}$, yields the transfield
force-balance equation (the generalized Grad-Shafranov
equation)
 \eqb
\frac 1{\cal
R}\left[\rho\gamma^2v_p^2+\frac{E^2-B_p^2}{4\pi}\right]-
\mathbf{\widehat n}\cdot\nabla\frac{B_p^2}{8\pi}+
\frac 1{r^2}\rho\gamma^2v_{\phi}^2\mathbf{\widehat n}\cdot\mathbf{r}=
\frac1{8\pi r^2}\mathbf{\widehat n}\cdot\nabla
\left[r^2(B^2_{\phi}-E^2)\right];
 \label{transfield}\eqe
where $\cal R$ is the local curvature radius of the poloidal
field line (defined such that $\cal R$ is positive when the flux surface is concave so that the collimation
angle decreases),
 \eqb
\frac 1{\cal R}= -\mathbf{\widehat{n}\cdot}(\mathbf{\widehat{l}\cdot\nabla})\mathbf{\widehat{l}}=
\mathbf{\widehat{n}\cdot}[\mathbf{\widehat{l}\times}(\mathbf{(\mathbf{\nabla\times\widehat{l}})})]=
-\mathbf{\widehat{\phi}\cdot}(\mathbf{\nabla\times\widehat{l}}).
 \eqe
Eqs. (\ref{EBp}), (\ref{Omega}), (\ref{continuity}), (\ref{energy}), (\ref{momentum})
and (\ref{transfield}) form a complete set of equation describing cold, axisymmetric MHD flows. This set
could be reduced to a pair of equations for $\Psi$ and $\gamma$.

Eliminating $B_{\phi}$ from Eqs. (\ref{energy}) and (\ref{momentum}), one
can express the azimuthal velocity via $\Psi$ and $\gamma$ as
 \eqb
 v_{\phi}=\frac 1{\Omega r}\left(1-\frac{\mu-\Omega l}{\gamma}\right).
 \label{vphi}\eqe
Assuming for simplicity that at the origin of the outflow, the
rotation velocity is well below the speed of light,
$\Omega r_{\rm in},v_{\phi,{\rm in}}\ll 1$, one reduces Eq. (\ref{vphi}) to the
form
 \eqb
v_{\phi}=\frac 1{\Omega r}\left(1-\frac{\gamma_{\rm
in}}{\gamma}\right);
 \label{v_phi}\eqe
where the index "in" is referred to the parameters of the injected
plasma. Substituting this relation into Eq. (\ref{Omega})
and eliminating $B_{\phi}$ with the aid of Eq. (\ref{energy}),
one gets the expression for the poloidal velocity
 \eqb
v_p=\frac{r^2\Omega^2B_p}{\eta(\mu-\gamma)}\left[1-\frac
1{\Omega^2 r^2}\left(1-\frac{\gamma_{\rm
in}}{\gamma}\right)\right].
 \label{v_p}\eqe
Now one can write the identity $v_p^2+v_{\phi}^2+\gamma^{-2}=1$ as the Bernoulli equation
 \eqb
\frac{\Omega^4r^4B_p^2}{\eta^2(\mu-\gamma)^2}
\left[1-\frac 1{\Omega^2 r^2}\left(1-\frac{\gamma_{\rm
in}}{\gamma}\right)\right]^2 +\frac 1{\Omega^2
r^2}\left(1-\frac{\gamma_{\rm in}}{\gamma}\right)^2+\frac
1{\gamma^2}=1;
 \label{Bernoulli}\eqe
which connects the Lorentz factor of the flow with the geometry of the flux tube defined by
the function $\Psi$.

The transfield equation (\ref{transfield}) is converted into an equation for $\Psi$ and $\gamma$
upon substituting
$E$ from Eq. (\ref{Efield}), $B_{\phi}$ from Eq. (\ref{energy}), $\rho$ from Eq. (\ref{continuity}), $v_p$
from Eq. (\ref{v_p}) and $v_{\phi}$ from Eq. (\ref{v_phi}). Therefore Eqs. (\ref{transfield}) and
(\ref{Bernoulli}) form a complete set of equations.

\section{Boundary conditions and integrals of motion}

At the inlet of the flow, one should specify the distribution of the poloidal flux or, which is the same,
of the poloidal magnetic field $B_p$. We are interested in outflows subtending a finite magnetic flux,
$\Psi_0$, therefore we have to prescribe a boundary condition at the last magnetic surface. If the flow
is confined by the pressure of the external medium, the pressure
balance condition should be satisfied at the boundary.
In the proper plasma frame, the magnetic field is
$B'=(B^2-E^2)^{1/2}=(B_{\phi}^2+(1-\Omega^2r^2)B_p^2)^{1/2}$. The condition that the pressure of this
field is compensated by the external pressure is written as
 \eqb
[B_{\phi}^2+(1-\Omega^2r^2)B_p^2]_{\Psi(r,z)=\Psi_0}=8\pi p_{\rm ext}(r,z);
 \label{bound}\eqe
where $p_{\rm ext}$ is the pressure of the external medium.

 In the cold flow, one has also to prescribe the functions $\Omega(\Psi)$,
$\eta(\Psi)$ and $\gamma_{\rm in}(\Psi)$ at the inlet of the
flow so that only two integrals of motion, $l(\Psi)$ and
$\mu(\Psi)$, remain unknown. Assuming that the rotation
velocity is non-relativistic at the origin of the flow, we have
eliminated the dependence on $l$ (see transition from Eq.
(\ref{vphi}) to Eq. (\ref{v_phi})). In the general case, the
integral $l$ may be expressed via $\gamma_{\rm in}(\Psi)$
$\Omega(\Psi)$, $\eta(\Psi)$, $\mu(\Psi)$ and $B_p$ at the
inlet of the flow making use of (\ref{Omega}), (\ref{energy})
and (\ref{momentum}).
So one has to find only the energy integral, $\mu$.
This integral is determined by the condition of the smooth
passage of the flow through the singular surfaces, Alfven and
modified fast magnetosonic \citep{li_chiueh_begelman92,tsinganos_etal96,bogovalov97,
vlahakis_etal00,vlahakis_konigl03a}.
In the Pointing dominated outflows, the
Alfven surface coincides with the light cylinder, $\Omega r=1$,
whereas the fast magnetosonic surface goes into the far zone
$\Omega r\gg 1$. Transition through the Alfven surface could be
studied in the force-free approximation, i.e. neglecting the
plasma energy and inertia.

The force-free limit of the transfield equation is obtained by taking $\rho=0$ in
Eq. (\ref{transfield}). Making use of Eq. (\ref{EBp}), one finds
 \eqb 
(\Omega^2r^2-1)\frac{B_p^2}{\cal R}+
\frac 12\mathbf{\widehat{n}\cdot\nabla}\left[\left(\Omega^2r^2-1\right)B_p^2\right]
=\frac 1{2r^2}\mathbf{\widehat{n}\cdot\nabla}\left(rB_{\phi}\right)^2-
\Omega^2B_p^2\mathbf{\widehat{n}\cdot r}.
 \label{forcefree}\eqe 
In the force-free limit, the energy equation (\ref{energy}) is reduced to the form
 \eqb
 rB_{\phi}=2I(\Psi);\quad 2I(\Psi)=\eta(\Psi)\mu(\Psi)/\Omega(\Psi);
 \eqe
which means that the current flows along the magnetic surfaces. Now the force-free balance
equation can be recast in the form of a second order elliptical equation for $\Psi$ \citep{okamoto74},
which is sometimes called the pulsar equation. By inspecting Eq. (\ref{forcefree}), one
can see that in the pulsar equation, the second derivatives are multiplied by $(\Omega^2r^2-1)$ so that
the equation is singular at the light surface. The condition of regularity at this surface enables one
to fix the poloidal current $I(\Psi)$ (e.g.,
\citet{fendt97,contopoulos_etal99,uzdensky04,uzdensky05,lovelace_etal06,timokhin06}). Then the energy integral
is found just adding the matter energy flux as $\mu=\gamma_{\rm in}+2\Omega I/\eta$. The first term here is
small in the Poynting dominated outflows however, one cannot neglect it close to the axis where the current
$I$ goes to zero ($I=\pi\int jrdr=(\pi/2)j(r=0)r^2\to 0$ as $r\to 0$).

Note that decreasing of the energy flux towards the axis is the
generic property of the Poynting dominated outflows because the
poloidal current, $I(\Psi)$, always goes to zero at $\Psi\to
0$. Such a "hollow cone" energy distribution accounts, in
particular, for a specific morphology of the inner Crab and
other pulsar wind nebulae (e.g., review by
\citet{kirk_etal07}). In any case, the exact shape of
$\mu(\Psi)$ depends on the geometry of the flow close to the
source. In this paper, we study the flow in the far zone
therefore we assume that this function is given together with
other integrals of motion.

We would like only to note that the function $\mu(\Psi)$ has a universal form close to the axis.
The poloidal field remains finite at the axis so that
Eq.(\ref{Bfield}) yields
 \eqb
\Psi=\frac 12r^2B_p(r=0,z); 
\quad \Psi\to 0.
 \eqe
Beyond the light surface, the magnetic field becomes predominantly toroidal
whereas the flow becomes predominantly poloidal (see the next section), therefore Eq.
(\ref{Omega}) yields $B_{\phi}\approx E=r\Omega B_p$.    Then the
second term in the energy equation is written, close to the axis, as $r\Omega
B_{\phi}/\eta=\Omega^2r^2B_p(r=0,z)/\eta=2[\Omega(0)]^2\Psi/\eta(0)$. So close to the
axis, the energy integral has the universal form
 \eqb
\mu(\Psi)=\gamma_{\rm in}(0)\left(1+\frac{\Psi}{\widetilde{\Psi}}\right);\quad
\Psi\to 0;
 \label{energy1}\eqe
where
 \eqb
\widetilde{\Psi}=\frac{\gamma_{\rm in}(0)\eta(0)}{2[\Omega(0)]^2}.
 \eqe
Note that the flow is Poynting dominated only at $\Psi\gg\widetilde{\Psi}$.

An important point is that in outflows with a constant angular velocity, one can assume for the estimates
that the energy integral is described by the linear function (\ref{energy1}) not only close to the axis
but across  the jet. Both an analytical solution for the paraboloidal flux surfaces
\citep{blandford76,beskin_nokhrina06} and numerical simulations
\citep{komissarov_etal07,komissarov_etal08,tchekhovskoy08} show that this is a good approximation for such jets.

One can also obtain a quite general estimate for the energy
integral taking into account that beyond the light cylinder,
Eq. (\ref{Omega}) yields $B_{\phi}\approx -\Omega rB_p$, which
simply means that each revolution of the source adds to the
wind one more magnetic loop. Then the second term in Eq.
(\ref{energy}) may be estimated as $(\Omega r)^2B_p/\eta$.
Making use of the estimate $\Psi\approx (1/2)r^2B_p$ (the
coefficient is exact when the poloidal field is homogeneous),
one finds finally
 \eqb
\mu(\Psi)\approx\gamma_{\rm
in}+\frac{2\Omega^2(\Psi)\Psi}{\eta(\Psi)}.
 \label{energy_approx}\eqe
This expression provides a rough estimate for the energy
integral for arbitrary $\Omega(\Psi)$ and $\eta(\Psi)$.

\section{The basic equations in the limit $\Omega r\gg 1$}

\subsection{Expansion in $1/r$.}
We are interested in outflows initially dominated by the
Poynting flux. In such outflows, the Alfvenic surface, where
$B_{\phi}\approx B_p$, nearly coincides with the light surface
$\Omega r=1$. In the far zone, $\Omega r\gg 1$, the toroidal
field decreases as $B_{\phi}\propto 1/r$ , see Eq.
(\ref{energy}).
The poloidal field decreases as $1/r^2$ therefore in the far
zone, the field is nearly toroidal. The flow in the far zone
becomes nearly radial because according to Eq.(\ref{v_phi}),
$v_{\phi}\propto 1/r$. In spite of this, one generally have to
retain the terms with $B_p$ and $v_{\phi}$ in the equations.
The physical reason is that the hoop stress is nearly
compensated by the electric force so that one cannot generally
neglect small pressure of the poloidal field. The formal reason
is that the leading order terms in Eqs. (\ref{Bernoulli}) and
(\ref{transfield}) are the same, which makes the system nearly
degenerate, so that one have to retain smaller order terms.

In the transfield equation (\ref{transfield}), the leading
order terms are those in the right-hand side because the terms
in the left-hand side are small either as $B_p/B_{\phi}$ or as
$r/\cal R$. In the Bernoulli equation (\ref{Bernoulli}), one
gets the leading order terms just neglecting the terms with
$1/r$ and $1/\gamma$. This yields
 \eqb
\eta(\mu-\gamma)=\Omega^2r^2B_p;
 \label{Bernoulli0}\eqe
or, according to Eqs. (\ref{EBp}) and
(\ref{energy}),
 \eqb B_{\phi}+E=0.
 \label{zeroth_order}\eqe
If one substituted this relation into the right-hand side of the
transfield equation, one would kill the leading order terms.
The correct procedure \citep{vlahakis04} is to expand the
Bernoulli equation (\ref{Bernoulli}) to the first non-vanishing
order in $1/r$ and $1/\gamma$ and only then to eliminate the
leading order terms from Eq.(\ref{transfield}). Expanding Eq.
(\ref{Bernoulli}) yields
 \eqb
B_{\phi}^2-E^2\equiv\left(\frac{\eta(\mu-\gamma)}{\Omega r}\right)^2-(\Omega r)^2B_p^2=
\left(\frac{\Omega^2r^2+\gamma_{\rm in}^2}{\gamma^2}
-1\right)B_p^2.
 \label{Bernoulli1}\eqe
Substituting this relation into the right-hand side of Eq.
(\ref{transfield}), one gets
 \eqb 
\frac 1{\cal
R}\left(\rho\gamma^2v_p^2+\frac{E^2-B_p^2}{4\pi}\right)+ \frac
1{r^2}\left(\frac{B_p^2}{4\pi}+
\rho\gamma^2v_{\phi}^2\right)\mathbf{\widehat n}\cdot\mathbf{r} 
=\frac1{8\pi r^2}\mathbf{\widehat n}\cdot\nabla
\left[\frac{\Omega^2r^4B^2_p}{\gamma^2}\left(1+\frac{\gamma_{\rm in}^2}{\Omega^2r^2}\right)\right].
 \label{tr}\eqe 
In this equation, there are no terms which nearly cancel each
other. Therefore one can now retain only terms of the lowest
order in $1/r$ and $1/\gamma$, e.g., neglecting $B_p$ with
respect to $E$ or substituting $v_p$ by unity. Moreover, one
can now use Eq.(\ref{zeroth_order}), which is the zeroth order
approximation to the Bernoulli equation, in order to further
simplify this equation. For example, the expression in the first
brackets in the left-hand side could be transformed, with the
aid of Eqs. (\ref{EBp}), (\ref{continuity}) and (\ref{energy}),
as
 \eqb
\rho\gamma^2v_p^2+\frac{E^2-B_p^2}{4\pi}=
\frac 1{4\pi}(4\pi \rho\gamma^2v_p-\Omega rB_pB_{\phi})=\frac{\mu\eta B_p}{4\pi}.
 \eqe
The expression in the second brackets in the left-hand side of
Eq.(\ref{tr}) could also be simplified in the same way after
substituting $v_{\phi}$ from Eq. (\ref{v_phi}):
 \begin{eqnarray}
\frac{B_p^2}{4\pi}+ \rho\gamma^2v_{\phi}^2=\frac 1{4\pi\Omega^2
r^2}\left[-\Omega
rB_pB_{\phi}+4\pi\rho\gamma^2v_p\left(1-\frac{\gamma_{\rm
in}}{\gamma}\right)^2\right] \\
=\frac{B_p}{4\pi\Omega^2
r^2}\left[-\Omega rB_{\phi}+\eta\gamma\left(1-\frac{\gamma_{\rm
in}}{\gamma}\right)^2\right]=\frac{\eta B_p}{4\pi\Omega^2
r^2}\left(\mu-2\gamma_{\rm in}+\frac{\gamma^2_{\rm
in}}{\gamma}\right).
 \nonumber\end{eqnarray}
We can also use Eq. (\ref{Bernoulli0}) in the right-hand side
of this equation. Eventually one finds
 \eqb 
\frac{\mu\eta B_p}{\cal R}+ \frac{\eta
B_p}{\Omega^2r^4}\left(\mu-2\gamma_{\rm in}+\frac{\gamma_{\rm
in}^2}{\gamma}
\right)\mathbf{\widehat n}\cdot\mathbf{r} 
=\frac1{2r^2}\mathbf{\widehat n}\cdot\nabla
\left[\frac{\eta^2(\mu-\gamma)^2}{\Omega^2\gamma^2}\left(1+\frac{\gamma_{\rm
in}^2}{\Omega^2r^2}\right)\right].
 \label{tr1}\eqe 
This is the asymptotic transfield equation valid at $\Omega r\gg 1$. It may be significantly simplified
in specific cases.

\subsection{Asymptotic transfield equation in different regimes.}
If the flow is initially Poynting dominated, one can neglect the terms with $\gamma_{\rm in}$ far enough from the axis, $\mu,\,\Omega r\gg\gamma_{\rm in}$. Then one comes to the
equation obtained, in a different form, by \citet{vlahakis04}:
   \eqb
\mu\eta B_p\left(\frac 1{\cal R}+ \frac
{\mathbf{\widehat{n}\cdot r}}{\Omega^2r^4}\right)= \frac1{2r^2}
\mathbf{\widehat
n}\cdot\nabla\frac{\eta^2(\mu-\gamma)^2}{\Omega^2\gamma^2}.
 \label{asympt_transfield1}\eqe
The solution to this equation describes the main body of the
flow but it could not be continued to the axis because it could
not satisfy the condition $\Psi(r=0)=0$. Close to the axis,
where the Poynting flux decreases according to Eq.
(\ref{energy1}), the terms with $\gamma_{\rm in}$ should be
retained therefore one should generally solve the full
asymptotic equation (\ref{tr1}).

Taking into account that the terms with $\gamma_{\rm in}$ play
role only close to the axis, where the angular velocity and the
injection Lorentz factor could be considered as constants, one
can present Eq. (\ref{tr1}), with the aid of Eq.
(\ref{Bernoulli0}), in a more convenient form
 \eqb 
\eta\mu B_p\left[\frac 1{\cal R}+\frac{\mathbf{\widehat
n}\cdot\mathbf{r}}{\Omega^2r^4}\left(1-\frac{2\gamma_{\rm
in}}{\mu}+\frac{\gamma_{\rm in}^2}{\gamma^2}\right) \right]
=\frac1{2r^2}\left(1+\frac{\gamma_{\rm
in}^2}{\Omega^2r^2}\right) \mathbf{\widehat
n}\cdot\nabla\frac{\eta^2(\mu-\gamma)^2}{\Omega^2\gamma^2}.
 \eqe 
The solution to this equation could be continued to the axis in
spite of the fact that the equation was formally derived at the
assumption $\Omega r\gg 1$. The reason is that far enough from
the center, the light cylinder $\Omega r=1$ is well within the
matter dominated zone, $\Psi\ll {\widetilde\Psi}$, where the
flow is practically hydrodynamic.

In the most interesting case of collimated flows, $z\gg r$, one
can take $\mathbf{\widehat n}\cdot\mathbf{r}=r$ and
$\mathbf{\widehat n}\cdot\nabla=\partial/\partial r$. When
looking for the shape of the magnetic surfaces, one can
conveniently use the unknown function $r(\Psi,z)$ instead of
$\Psi(r,z)$. Then, e.g.,
 \eqb
B_p=\frac 1r\vert\nabla\Psi\vert\approx\frac 1r\frac{\partial\Psi}{\partial r}=
\left(r\frac{\partial r}{\partial\Psi}\right)^{-1}.
 \label{Bp_jet}\eqe
In the same approximation, the curvature radius may be presented as
(note that $\cal R$ is defined to be positive for concave surfaces)
 \eqb
\frac 1{\cal R}=-\frac{\partial^2r}{\partial z^2}.
 \label{curvature}\eqe
Now the transfield equation for the collimated flows in the far
zone could be written as
 \eqb 
\eta\mu\left[-\frac{\partial^2r}{\partial
z^2}+\frac{1}{\Omega^2r^3}\left(1-\frac{2\gamma_{\rm
in}}{\mu}+\frac{\gamma_{\rm in}^2}{\gamma^2}\right) \right]
=\frac1{2r}\left(1+\frac{\gamma_{\rm in}^2}{\Omega^2r^2}\right)
\frac{\partial}{\partial\Psi}\frac{\eta^2(\mu-\gamma)^2}{\Omega^2\gamma^2}.
\label{asympt_transfield}\eqe 
We believe that this equation suits well to numerical solution
because it does not contain terms that nearly cancel each
other. In many cases it could even be solved analytically. For
analytical solution, this equation could be conveniently
considered in two overlapping domains, namely, in the main body
of the jet, where the flow is significantly accelerated so that
one can neglect the terms with $\gamma_{\rm in}/\gamma$, and
close to the axis, where the flux surfaces are nearly straight
so that one can neglect the curvature term
$\partial^2r/\partial z^2$. Solutions in these domains are
smoothly matched in the intermediate zone.

Close to the axis, where the flux surfaces are nearly straight,
 \eqb
 \frac{d^2r}{dz^2}\ll \frac 1{\Omega^2r^3}
  \label{equilibrium_cond}\eqe
one can neglect the term with the derivative in $z$ and write
the transfield equation as an ordinary differential equation
(see also \citet{beskin_malyshkin00,beskin_nokhrina06,beskin_nokhrina08})
 \eqb
\mu\left(1+\frac{\gamma_{\rm
in}^2}{\gamma^2}\right)-2\gamma_{\rm in}=
\frac{\Omega^2r^2+\gamma_{\rm in}^2}{\Omega\gamma}(\mu-\gamma)
\frac{\partial}{\partial\Psi}\frac{\eta(\mu-\gamma)}{\Omega\gamma}.
 \label{axis_transfield}\eqe
We will analyze it in sect. 8. In some cases, the condition
(\ref{equilibrium_cond}) is fulfilled across the whole jet;
then the full jet structure is described by the one-dimensional
equation, the $z$ dependence entering only via the boundary
conditions.

Note that neglecting the derivative in $z$ in the the
transfield equation, one comes to the equation describing
cylindrical equilibria.  We will refer to such a situation as
an {\it equilibrium collimation} in the sense that at any $z$,
the jet structure is the same as in an appropriate equilibrium
cylindrical configuration.

At $\Psi\gg\widetilde{\Psi}$, the plasma is significantly accelerated in the far zone so that the transfield
equation is reduced to:
 \eqb
2\mu\eta r\left(-\frac{\partial^2r}{\partial z^2} + \frac
1{\Omega^2r^3}\right) =
\frac{\partial}{\partial\Psi}\frac{\eta^2(\mu-\gamma)^2}{\Omega^2\gamma^2}.
 \label{Poynting_jet}\eqe
This equation describes the structure of the main body of the jet. One cannot give a simple physical interpretation of terms in this equation however, one can gain some physical insight
considering regimes when different terms dominate. If the condition (\ref{equilibrium_cond})
is fulfilled across the jet, one can neglect the derivative in $z$ thus coming to a $\Psi\gg\widetilde{\Psi}$
limit of Eq. (\ref{axis_transfield}). In this case, the jet as a whole is collimated
in the equilibrium regime. On the other hand, in some configurations (and anyway far enough from
the axis) the condition opposite to (\ref{equilibrium_cond}) is fulfilled; then the term with
the second derivative becomes dominant so that the equation is
reduced to
 \eqb
-2\mu\eta r\frac{\partial^2r}{\partial z^2}=
\frac{\partial}{\partial\Psi}\frac{\eta^2(\mu-\gamma)^2}{\Omega^2\gamma^2}.
 \label{nonequilibrium}\eqe
This equation could be directly obtained assuming that the
field is purely toroidal whereas the flow is purely poloidal
\citep{lyubarsky_eichler01}. Then the flux freezing condition (\ref{freezing}) yields
$B_{\phi}^2-E^2=(B_{\phi}/\gamma)^2$. Substituting this relation into the transfield equation (\ref{transfield})
and dropping the terms with $B_p$ and $v_{\phi}$, one comes, in the far zone, to Eq. (\ref{nonequilibrium}). In this case, the function $\Psi$ could be considered as a specially
normalized stream function, $\rho\gamma\mathbf{v}\propto (1/r)\nabla\Psi\times{\tilde\phi}$,
which describes the flow lines. The
poloidal field, $B_p$, and the angular velocity, $\Omega$, become just auxiliary quantities formally
defined by Eqs. (\ref{Bfield}) and (\ref{Efield}). Note that Eq. (\ref{nonequilibrium}) does not change under the transformation $\Psi\to a\Psi$; $\Omega\to a^{-1}\Omega$;
$\eta\to a^{-1}\eta$; where $a$ is an arbitrary number. As it follows from Eqs. (\ref{Bfield}), (\ref{Efield}) and (\ref{continuity}),
neither poloidal velocity nor the electric field change under this transformation
whereas $B_p$ and $\Omega$ could acquire any values.
The flow with the poloidal field and azimuthal velocity
neglected could be seen as composed from coaxial magnetic loops
moving away and expanding together with the plasma. In this
case, the difference between the hoop stress and the electric
force is not counterbalanced by the pressure of the poloidal
field. Taking into account that the electric field could not
compensate the hoop stress completely (in the frame moving with
the loop, the $r$ component of the electric field is zero), one
concludes that there is a residual force towards the axis of the flow.
This does not
mean that the flow immediately converges to the axis because in
highly relativistic flows, the residual of the hoop stress and
the electric force is small. 
In any case, we will refer to this situation as a {\it
non-equilibrium collimation}. We will see that different regimes
of collimation correspond to different acceleration regimes.

\subsection{Asymptotic Bernoulli equation and boundary conditions}

The transfield equation should be supplemented by the Bernoulli
equation. We have already used this equation in the zeroth
order in $1/r$, Eq. (\ref{Bernoulli0}), when simplified the
transfield equation. However, one should be careful when using
this equation in order to find $\gamma$ because $\gamma$ turns
out to be a small difference between two large terms if the flow is
Poynting dominated, $\mu\gg\gamma$. Therefore
$\gamma$ could be found from the Bernoulli equation in the form
of Eq. (\ref{Bernoulli0}) only if $\sigma$ is not too large.
Generally one should retain higher order terms
and use, e.g., Eq. (\ref{Bernoulli1}). Without loss of
accuracy, this equation could be written as a cubic equation
for $\gamma$ (e.g. \citet{beskin_etal98})
 \eqb
\mu-\frac{\Omega^2r^2B_p}{\eta}-\gamma=\frac{\Omega^2r^2B_p}{2\gamma^2\eta}\left(1-\frac{\gamma^2-\gamma^2_{\rm
in}}{\Omega^2r^2}\right).
 \label{cubic}\eqe
This equation is reduced to the zeroth order Bernoulli equation
(\ref{Bernoulli0}) if one could neglect the expression in the
right-hand side. This expression is small as compared with
$\mu$ however, it could be neglected only when it is less than
$\gamma$, i.e. only if $\gamma^3\gg\mu$; $(\Omega r)^3\gg\mu$.
Note that $\gamma=\mu^{1/3}$ when the flow velocity is equal to
the fast magnetosonic velocity (e.g., \citet{camenzind86}) so
that one can find $\gamma$ from the zeroth order Bernoulli
equation only beyond the fast magnetosonic point. Well within
this point, the Lorentz factor could be found from another
limit of Eq. (\ref{cubic}):
 \eqb
\left(1-\frac{\Omega^2r^2B_p}{\mu\eta}+\frac
1{2\Omega^2r^2}\right)\gamma^2=\frac
12\left(1+\frac{\gamma_{\rm in}}{\Omega^2r^2}\right).
 \eqe
Generally one has to solve the cubic equation (\ref{cubic}) so
that there is no simple expression for $\gamma$ valid in the
whole far zone. This in fact means that the acceleration
regimes inside and outside the fast magnetosonic point could be
different. For example, in the split monopole wind, which
represents a non-confined wind from a point source, the Lorentz
factor grows linearly with the radius until the fast
magnetosonic point and then the acceleration becomes
logarithmically slow \citep{beskin_etal98}.

Note that the asymptotic transfield equation
(\ref{asympt_transfield}) is valid both outside and inside the
fast magnetosonic point because it was derived only under the
condition $\Omega r\gg 1$. In the next section, we show that
when considering collimated, Poynting dominated flows, one can
avoid finding $\gamma$ from the Bernoulli equation. In this
case, the acceleration regime does not change at the fast
magnetosonic point therefore this point will not appear more in
this paper.

The asymptotic form of the boundary condition (\ref{bound}) may
be found by making use of Eq.(\ref{Bernoulli1}) and taking into
account that $\gamma\gg\gamma_{\rm in}$ in the outer parts of
the Poynting dominated jet; this yields
 \eqb
\left(\frac{\Omega
rB_p}{\gamma}\right)^2_{\Psi(r,z)=\Psi_0}=8\pi p_{\rm ext}(z).
 \eqe
Taking into account Eq. (\ref{Bernoulli0}), one could write
this condition also as
 \eqb
\left(\frac{\eta(\mu-\gamma)}{\Omega
r\gamma}\right)^2_{\Psi(r,z)=\Psi_0}=8\pi p_{\rm ext}(z).
 \label{boundary}\eqe

\section{The Poynting dominated flow in the far zone}

Let us first consider the structure of the Poynting dominated
flow, $\mu\gg \gamma$. Since the Poynting flux goes to zero at
the axis, (see Eq. (\ref{energy1})), this approximation is
violated close enough to the axis,
$\Psi\lesssim\widetilde{\Psi}$. Moreover, we will find that the
flow is accelerated in such a way that the closer the field
line to the axis, the earlier (at a smaller $z$) the flow
kinetic energy approaches the total energy. Therefore a
$\sigma\sim 1$ core is anyway presented within the Poynting
dominated jet so that the results of this section could not be
applied close enough to the axis. In Section 8, we find the
structure of the flow close to the axis, which is smoothly
matched, at a larger $r$, with the solution for the Poynting
dominated flow.

\subsection{The governing equation}

Here we study the structure of the flow at
$\Psi\gg\widetilde{\Psi}$, i.e., when the Poynting flux
initially exceeded the plasma kinetic energy. In this case,
we can use the asymptotic transfield equation in the form (\ref{Poynting_jet}).

As it was discussed in sect. 4.3, one cannot find a simple
expression for $\gamma$ from the Bernoulli equation in order to
substitute it into the transfield equation and obtain a single
equation for $\Psi$. On the order hand, $\gamma$ could be
easily found from the transfield equation provided the shape of
the magnetic surfaces, $r(\Psi,z)$, is known. An important
point is that in this case, an extra accuracy is generally not
necessary because in the transfield equation, $\gamma$ is not
presented as a difference of large terms. A special care should
be taken only if the flow becomes nearly radial because the
curvature of the flux surfaces is determined in this case by
small deviations of the flow lines from the straight lines;
this case will be specially addresses in sect. 7.2. In this and
the next sections, we will neglect corrections of the order of
$\gamma/\mu$ to the shape of the flux line; then the Bernoulli
equation (\ref{Bernoulli0}) is reduced to
 \eqb
\Omega^2r^2B_p=\eta\mu;
 \label{Bernoulli0a}\eqe
which could be considered, with
account of Eq.(\ref{Bp_jet}), as an equation for $r(\Psi,z)$:
 \eqb
\mu\eta \frac{\partial r}{\partial\Psi}=r\Omega^2.
 \eqe
The solution to this equation is presented as
 \eqb
r=D(z)\Phi(\Psi);\quad
\Phi(\Psi)=\sqrt{2}\exp\left(\int_{\widetilde{\Psi}}^{\Psi}\frac{\Omega^2d\Psi}{\mu\eta}\right);
 \label{x(psi}\eqe
where $D(z)$ is an arbitrary function. One sees that the structure of collimated, Poynting dominated
jets is generally self-similar. Recall that this
equation is valid only at $\Psi\gg\widetilde{\Psi}$; the
solution will be continued to the axis in the Section 8.
In any case, $D$ is roughly the radius of the very inner
part of the jet, $\Psi\sim\widetilde{\Psi}$.

Close enough to the axis, one can use Eq. (\ref{energy1}) for
$\mu$, which implies
 \eqb
\Phi=\sqrt{1+\frac{\Psi}{\widetilde{\Psi}}}. 
 \label{Phi}\eqe
This means the poloidal magnetic field becomes homogeneous,
$\Psi\propto r^2$; $\partial B_p/\partial r=0$,well inside the
jet, ${\tilde\Psi}\ll\Psi\ll\Psi_0$. Note that when finding the expression (\ref{energy1}) for
$\mu$, we assumed that the poloidal field is homogeneous near
the axis so that this result is nothing more than a consistency
check. The same expression for $\Phi(\Psi)$ is also obtained if
one uses the general estimate (\ref{energy_approx}) for $\mu$.
This is also because the coefficient 2 in (\ref{energy_approx})
corresponds to the homogeneous poloidal field. Another
coefficient would result in a power law function $\Phi(\Psi)$.
Such a strong dependence on $\mu$ arises because $\mu$ enters in
the exponent.

In order to find the function $D(z)$, let us substitute
Eq.(\ref{x(psi}) into the left-hand side of Eq. (\ref{Poynting_jet}) and integrate
the obtained equation between $\widetilde{\Psi}$ and $\Psi_0$:
 \eqb
-2D\frac{d^2D}{dz^2}\int_{\widetilde{\Psi}}^{\Psi_0}\Phi^2\mu\eta
d\Psi +\frac
2{D^2}\int_{\widetilde{\Psi}}^{\Psi_0}\frac{\mu\eta
d\Psi}{\Omega^2\Phi^2}=\left(\frac{\eta\mu}{\Omega\gamma}\right)^2_{\Psi=\Psi_0}-
\left(\frac{\eta\mu}{\Omega\gamma}\right)^2_{\Psi=\widetilde{\Psi}}.
 \eqe
Note that the region $\Psi\sim\Psi_0$ contributes mostly into
the integrals therefore we could choose $\widetilde{\Psi}$ as
the lower limit of integration even though the solution
(\ref{x(psi}) is no longer valid there. One can also neglect
the last term in the right hand side as compared with the first
one because it could be checked a posteriori that the
expression in the brackets grows with $r$. Making use of the
boundary condition (\ref{boundary}), one reduces the right-hand
side of this equation to $8\pi r^2p_{\rm ext}=4\pi
\Phi^2D^2p_{\rm ext}$. Then one gets the equation for $D(z)$ in
the closed for
 \eqb
\frac{d^2D}{dz^2}\int_{\widetilde{\Psi}}^{\Psi_0}\Phi^2\mu\eta
d\Psi -\frac 1{D^3}\int_{\widetilde{\Psi}}^{\Psi_0}\frac{\mu\eta
d\Psi}{\Omega^2\Phi^2}=-4\pi \left[\Phi(\Psi_0)\right]^2p(z)D.
 \label{D(Z)}\eqe
This equation could be written in the dimensionless form as
 \eqb
\frac{d^2Y}{dZ^2}-\frac 1{Y^3}+\beta {\cal P}(Z)Y=0;
 \label{Y(Z)}\eqe
where \eqb Z=\Omega(\Psi_0)z;\quad
Y(z)=\alpha^{-1/4}\Omega(\Psi_0)D(z);
 \label{Y}\eqe
 \eqb
\alpha=[\Omega(\Psi_0)]^2\left(\int_{\widetilde{\Psi}}^{\Psi_0}\frac{\mu\eta
d\Psi}{\Omega^2\Phi^2}\right)\left(\int_{\widetilde{\Psi}}^{\Psi_0}\Phi^2\mu\eta
d\Psi\right)^{-1}; \label{alpha}\eqe\eqb \beta=4\pi
p_0\left[\frac{\Phi(\Psi_0)}{\Omega(\Psi_0)}\right]^2
\left(\int_{\Psi_p}^{\Psi_0}\Phi^2\mu\eta d\Psi\right)^{-1}.
 \label{beta}\eqe
 \eqb
p_0=p_{ext}\left(z=1/\Omega(\Psi_0)\right);\quad p_0{\cal P}(Z)=p(z).
 \eqe
This equation generalizes the equation for the jet radius
obtained by \citet{komissarov_etal08} as an order of magnitude
estimate. We see that this equation is in fact asymptotically
exact. Moreover, finding $Y(Z)$ from this equation, one finds
the full structure of the flow. Therefore we will call
Eq.(\ref{Y(Z)}) the governing equation for Poynting dominated
jets.

Note that there is one to one correspondence between the terms
in the governing equation and in the original asymptotic
transfield equation (\ref{Poynting_jet}). Namely the pressure
term (the last one) in Eq. (\ref{Y(Z)}) comes from the
right-hand side of Eq. (\ref{Poynting_jet}) whereas the first
two terms correspond to the terms in the left-hand side.
Following the discussion in sect. 4.2, one sees that the
collimation is in the equilibrium regime if the second term
dominates the first one. Then one immediately finds
$Y(Z)=\left[\beta{\cal P}(Z)\right]^{-4}$. Of course neglecting
the derivative in the equation, one looses solutions. The lost solutions just describe oscillations with
respect to the equilibrium state. If the jet is not very
narrow, the term $Y^{-3}$ becomes negligibly small; then the
governing equation becomes linear. In this case the jet is
collimated in the non-equilibrium regime. In Sect. 6. we
present a more detailed analysis for the case of a power law
profile of the external pressure.

The solution to the governing equation (\ref{Y(Z)}) could be
presented \citep{polyanin_zaitsev02} in the form $Y=wy$, where
the auxiliary function, $w$, satisfies the linear equation
 \eqb
\frac{d^2w}{dZ^2}+\beta{\cal P}(Z)w=0.
 \label{lin_app}\eqe
Then the equation for $y$ has the first integral
 \eqb
\left(w^2\frac{dy}{dZ}\right)^2=C_1-\frac 1{y^2};
 \eqe
which could be immediately integrated once again. Now the general
solution to Eq. (\ref{Y(Z)}) is found as
 \eqb
 Y=w\left[\frac 1{C_1}+C_1\left(C_2+\int\frac{dZ}{w^2}\right)^2\right]^{1/2}.
 \label{solution}\eqe
In section 6, we
present such a solution for the jet with a constant angular
velocity confined by the external pressure decreasing as a
power law.

\subsection{Finding the structure of the flow}

According to Eq. (\ref{x(psi}), the flux surfaces are self-similar in the Poynting dominated domain.
Having found $Y$ from the governing equation, one finds the shape of the magnetic surfaces as
 \eqb
\Omega(\Psi_0)r(\Psi,Z)=\alpha^{1/4}\Phi(\Psi)Y(Z).
 \label{jet_radius}\eqe
Taking into account that the region $\Psi\sim\Psi_0$
contributes mostly into the integrals in Eqs. (\ref{alpha}) and
(\ref{beta}), one can estimate the coefficients in the governing
equation as
 \eqb
\alpha\sim[\Phi(\Psi_0)]^{-4};\quad\beta\sim \frac{2\pi
p_0}{[\Omega(\Psi_0)]^4\Psi_0^2}.
 \eqe
In the last equality, we used the estimate
(\ref{energy_approx}). Substituting the obtained estimate for
$\alpha$ into Eq. (\ref{jet_radius}), one finds
 \eqb
 Y\sim r(\Psi_0)\Omega(\Psi_0);
 \eqe
so that $Y$ is of the order of the dimensionless outer radius
of the jet. Making use of the estimate $\Psi\sim (1/2)r^2B_p$,
one can write
 \eqb
\beta\sim\left(\frac{8\pi
p_0}{B^2}\right)_{\Omega(\Psi_0)r(\Psi_0)=1};
 \eqe
so that $\beta$ is of the order of the ratio of the external
pressure to the magnetic pressure at the base of the flow.

In order to find the Lorentz factor of the flow, one substitutes
Eq. (\ref{jet_radius}) into the left hand side of Eq.
(\ref{Poynting_jet}) and performs integration between
$\widetilde{\Psi}$ and $\Psi$ to obtain
 \eqb
-2\sqrt{\alpha}Y\frac{d^2Y}{dZ^2}\int_{\widetilde{\Psi}}^{\Psi}\Phi^2\mu\eta
d\Psi
+\frac{2\Omega^2(\Psi_0)}{\sqrt{\alpha}Y^2}\int_{\widetilde{\Psi}}^{\Psi}\frac{\mu\eta
d\Psi}{\Omega^2\Phi^2}=\left(\frac{\eta\mu}{\Omega\gamma}\right)^2_{\Psi=\Psi}-
\left(\frac{\eta\mu}{\Omega\gamma}\right)^2_{\Psi=\widetilde{\Psi}}.
 \eqe
Retaining only the first term in the right-hand side, as it was done in Eq. (\ref{D(Z)}), one gets
the relation for $\gamma(\Psi,Z)$ in the closed form
 \eqb
 \left(\frac{\eta\mu}{\Omega\gamma}\right)^2=
-2\sqrt{\alpha}Y\frac{d^2Y}{dZ^2}\int_{\widetilde{\Psi}}^{\Psi}\Phi^2\mu\eta
d\Psi
+\frac{2\Omega^2(\Psi_0)}{\sqrt{\alpha}Y^2}\int_{\widetilde{\Psi}}^{\Psi}\frac{\mu\eta
d\Psi}{\Omega^2\Phi^2}.
 \label{Lorentz-factor}\eqe
Specifically for the periphery of the flow, $\Psi=\Psi_0$, one
finds, with the aid of Eqs. (\ref{Y(Z)}) and (\ref{alpha}),
 \eqb
\gamma(\Psi_0,Z)=\frac W{\sqrt{\beta {\cal P}(Z)}Y(Z)};
\quad W=\frac{\eta(\Psi_0)\mu(\Psi_0)}{\sqrt{2[\Omega(\Psi_0)]^3}}
\left(\int_{\widetilde{\Psi}}^{\Psi_0}\frac{\mu\eta
d\Psi}{\Omega^2\Phi^2}\int_{\widetilde{\Psi}}^{\Psi_0}\Phi^2\mu\eta
d\Psi\right)^{-1/4}.
 \eqe

With the aid of Eq. (\ref{curvature}), one can write Eq. (\ref{Lorentz-factor}) as
 \eqb
\frac 1{\gamma^2}=A\frac r{\cal R}+B\frac 1{\Omega^2r^2};
 \label{gamma_general}\eqe
where
 \eqb
A(\Psi)=2\left[\frac{\Omega(\Psi)}{\Phi(\Psi)\eta(\Psi)\mu(\Psi)}\right]^2
\int_{\widetilde{\Psi}}^{\Psi}\Phi^2\mu\eta d\Psi;
 \eqe\eqb
 B(\Psi)=2\left[\frac{\Omega^2(\Psi)\Phi(\Psi)}{\eta(\Psi)\mu(\Psi)}\right]^2
 \int_{\widetilde{\Psi}}^{\Psi}\frac{\mu\eta
d\Psi}{\Omega^2\Phi^2}.
 \eqe
This equation generalizes the equation obtained by
\citet{tchekhovskoy08} and by \citet{komissarov_etal08}. Making
use of the estimate (\ref{energy_approx}) for $\mu$, one finds
that the coefficients $A$ and $B$ are always of the order of
unity. In the case of equilibrium collimation, when the
condition (\ref{equilibrium_cond}) is fulfilled, one can
neglect the first term in the right-hand side, which yields the
old-established \citep{buckley77} acceleration regime
$\gamma\propto\Omega r$. In the opposite limit of
non-equilibrium collimation, one comes to the scaling
$\gamma\propto\sqrt{{\cal R}/r}$ found recently by
\citet{beskin_etal04}.

\section{Poynting dominated jet with a constant angular velocity}
In this section, we apply the above general method to jets
with the constant angular velocity, $\Omega(\Psi)=\it const$.
In this case, one can conveniently use the dimensionless variables
 \eqb
X=\Omega r;\qquad Z=\Omega z.
 \eqe
We also assume that the injection is homogeneous, $\eta(\Psi)=\it const$.

Note that in this case, one can get simple relations assuming  that the
energy integral, $\mu$, is described by the linear function
(\ref{energy1}) not only close to the axis but across  the jet.
Then the poloidal flux is homogeneous, see Eq. (\ref{Phi}). The
coefficients $\alpha$ and $\beta$ defined by Eqs. (\ref{alpha})
and (\ref{beta}), correspondingly, are reduced to:
 \eqb
\alpha=3\left(\frac{\widetilde{\Psi}}{\Psi_0}\right)^2;
 \label{alpha_lin}\eqe
 \eqb
\beta=\frac{6\pi p_0}{\Omega^4\Psi^2_0}=\frac{6\pi p_0}{B_0^2};
 \label{beta_lin}\eqe
where 
$B_0\equiv \Omega^2\Psi_0$ is the characteristic magnetic field
at the light surface. Now the flux surfaces are described by a
simple formula
 \eqb
X= 3^{1/4}\left(\frac{\Psi}{\Psi_0}\right)^{1/2}Y(Z).
 \label{r_linear}\eqe
The coefficients in Eq. (\ref{gamma_general}) are reduced to
$A=1/3$; $B=1$ so that one gets the equation obtained by
\citet{tchekhovskoy08}. In terms of $Y$ and $\Psi$, the
equation for the Lorentz factor (\ref{Lorentz-factor}) could
now be written in the simple form
 \eqb
\frac{\sqrt{3}}{\gamma^2}=-\frac{\Psi}{\Psi_0}Y\frac{d^2Y}{dZ^2}+\frac{\Psi_0}{\Psi}\frac
1{Y^2}.
 \label{Lorentz_lin}\eqe
One sees that close enough to the axis, the second term in the
right-hand side dominates, which yields the acceleration regime
 \eqb
 \gamma=X.
 \label{acceleration1}\eqe
If $Y\gg \sqrt{Z}$, so that if collimation is not very strong,
the first term could dominate in the main body of the jet,
$\Psi\sim\Psi_0$. Then the Lorentz factor is determined by the
curvature of the magnetic surface, namely
 \eqb
\gamma=\sqrt{3}\left(X\frac{d^2X}{dZ^2}\right)^{-1/2}=\sqrt{3{\cal R}/r}.
 \label{acceleration2}\eqe
In any case, at $\Psi=\Psi_0$ one finds, with the aid of Eq.
(\ref{Y(Z)}),
 \eqb
\gamma(\Psi_0,Z)=\frac{3^{1/4}}{\sqrt{\beta {\cal P}(Z)}Y(Z)}.
 \label{gamma_p}\eqe
Below we present not only general formulae for the parameters of the flow
but also simple estimates with the aid of Eqs. (\ref{alpha_lin}), (\ref{r_linear}) and
(\ref{Lorentz_lin}).

Let the external pressure be decreasing as
 \eqb
{\cal P}=\frac 1{Z^{\kappa}}.
 \label{pressure}\eqe
Then the auxiliary equation (\ref{lin_app}) is solved via the Bessel functions so that
the general solution to the governing equation (\ref{Y(Z)}) could be found analytically.
Taking into account that the governing equation
is valid only at large $Z$, one can use only the appropriate
asymtotics of the solution. Since the
asymptotics depends on the sign of $\kappa-2$, let as consider
different cases separately.

 \subsection{The case $\kappa<2$}

In this case, a solution to Eq.
(\ref{lin_app}) is presented as
 \eqb
w=\sqrt{Z}J_{\frac 1{2-\kappa}}\left(\frac{2\sqrt{\beta}}{2-\kappa}Z^{1-\kappa/2}\right).
 \eqe
At a large $Z$, this function is reduced to
 \eqb
w=\sqrt{\frac{2-\kappa}{\pi}}\left(\frac{Z^{\kappa}}{\beta}\right)^{1/4}\cos S;\qquad
S=\frac{2\sqrt{\beta}}{2-\kappa}Z^{1-\kappa/2}-\frac{4-\kappa}{2-\kappa}\frac{\pi}4.
\eqe
Substituting this auxiliary function into Eq.
(\ref{solution}) yields the general solution to the governing equation
 \eqb
Y=\sqrt{\frac{2-\kappa}{\pi}}\left(\frac{Z^{\kappa}}{\beta}\right)^{1/4}\left[\frac
1{C_1}\cos^2S+C_1\left(C_2\cos S+ \frac{\pi}{2-\kappa}\sin S\right)^2\right]^{1/2}
 \label{Ykappa<2_gen} \eqe
At $C_1=(2-\kappa)/\pi$, $C_2=0$, this solution is reduced to a power law
 \eqb
Y=\left(\frac{Z^{\kappa}}{\beta}\right)^{1/4};
 \label{Ykappa<2}\eqe
which could be found directly from Eq. (\ref{Y(Z)}) by
neglecting the first term \citep{komissarov_etal08}. One sees
that with this solution, the first term in Eq.(\ref{Y(Z)}) is
much less, at $\kappa<2$, than the second one  so that the
collimation occurs in the equilibrium regime. The general
solution (\ref{Ykappa<2_gen}) also expands as $Z^{\kappa/4}$
but very long wave oscillations are superimposed on this
expansion, which means that the flow could oscillate around the
equilibrium state. Such oscillations are possible if the jet
was injected not in the equilibrium state. The amplitude of
these oscillations could be found by matching to the near zone
solution at $Z\sim 1$. The spatial period of these oscillations
increases with the distance as $Z^{\kappa/2}$.


Taking into account that the governing equation becomes algebraic in the equilibrium regime,
one can generalize Eq. (\ref{Ykappa<2}) to the general pressure distribution
provided the pressure decreases not faster than $z^{-2}$:
 \eqb
Y=\left(\beta\cal P\right)^{-1/4}.
 \label{Ygeneral}\eqe
One can see that the jet expands while the confining pressure decreases. When the jet
eventually enters the region with the constant pressure, the jet becomes cylindrical.

The Lorentz factor of the flow is found from Eq.
(\ref{Lorentz-factor}). For the smooth expansion described by
Eq. (\ref{Ykappa<2}) one can neglect the first term in the
right-hand side, which yields
 \eqb
\gamma=\frac{\eta\mu
Z^{\kappa/4}}{\Omega^2}\left(\frac{\alpha}{\beta}\right)^{1/4}
\left(2\int_{\widetilde\Psi}^{\Psi}\frac{\eta\mu
d\Psi}{\Phi^2}\right)^{-1/2}.
 \label{gamma_kappa<2}\eqe
One sees that in accord with the general analysis in section
5.2, the Lorentz factor of the flow is proportional to the
cylindrical radius, $\gamma\propto X$. If the energy integral
is a linear function of $\Psi$, Eq. (\ref{energy1}), which is
anyway the case well within the jet, this expression is reduced
just to $\gamma=X$, see Eq. (\ref{Lorentz_lin}). This estimate remains valid
also for non-power law pressure distributions when the jet shape is described by
Eq. (\ref{Ygeneral}). If the jet oscillates
with respect to the equilibrium expansion, as is described by Eq.
(\ref{Ykappa<2_gen}), the Lorentz factor also
oscillates with respect to that given by Eq.
(\ref{gamma_kappa<2}).

\subsection{The case $\kappa=2$}

In this case, a solution to Eq.
(\ref{lin_app}) is
 \eqb
w=\left\{\begin{array}{lll}
\sqrt{Z}\cos S; & S=\sqrt{\beta-1/4}\ln Z; & \beta>1/4;
\\ Z^{(1+\sqrt{1-4\beta})/2}; & & \beta<1/4.\end{array}\right.
 \eqe
Now the general solution to the governing equation is
 \eqb
 Y=\frac
1{\sqrt{C_1}}Z^{1/2}\left\{\begin{array}{ll}\left[\cos^2S+C_1^2\left(C_2\cos S+ \frac
1{\sqrt{\beta-1/4}}\sin S\right)^2\right]^{1/2}
& \beta>1/4; \\
Z^{(1/2)\sqrt{1-4\beta}}\left[1+
C_1^2\left(C_2-\frac
1{\sqrt{1-4\beta}Z^{\sqrt{1-4\beta}}}\right)^2\right]^{1/2};
& \beta<1/4.\end{array}\right.
 \eqe

The $\beta>1/4$ solution is similar to that for the $\kappa<2$
case. At specially chosen constants, $C_1=(\beta-1/4)^{1/2}$,
$C_2=0$, it is reduced to a pure power law
\citep{komissarov_etal08},
 \eqb
Y=\frac{Z^{1/2}}{(\beta-1/4)^{1/4}};
 \label{Ykappa=2a}\eqe
whereas generally long wavelength oscillations are superimposed
on the overall expansion. With the solution (\ref{Ykappa=2a}),
both terms in the left-hand side of the governing equation are
comparable so that this case is an intermediate between the
equilibrium and non-equilibrium collimation.

At $\beta<1/4$, the solution is reduced, at large $Z$, to a power law
\citep{komissarov_etal08}
 \eqb
Y=CZ^k;\quad k=(1+\sqrt{1-4\beta})/2.
 \label{Ykappa=2b}\eqe
Note that $1/2<k<1$ so that the flow is collimated but slower than in the case $\beta>1/4$.
The constant $C$ in this solution is not defined; it could be found
only by matching to the near zone solution. If the flow was not
collimated at $Z\sim 1$, there should be $C\sim 1$. This solution could be obtained directly
by neglecting the second term in the governing equation (\ref{Y(Z)}). This means that at $\beta<1/4$,
the collimation is non-equilibrium.


The Lorentz factor of the flow is found from Eq.
(\ref{Lorentz-factor}). At $\beta>1/4$, one substitutes the
solution (\ref{Ykappa=2a}), which yields the relation
 \eqb
\frac{\eta^2\mu^2Z}{2\Omega^2\gamma^2}=\frac
14\left(\frac{\alpha}{\beta-1/4}\right)^{1/2}
\int_{\widetilde{\Psi}}^{\Psi}\eta\mu\Phi^2 d\Psi+
\left(\frac{\beta-1/4}{\alpha}\right)^{1/2}
\int_{\widetilde{\Psi}}^{\Psi}\frac{\eta\mu}{\Phi^2}d\Psi;
 \eqe
which yields $\gamma\propto\sqrt{Z}\propto X$.
In this relation, the terms in the right-hand side are
comparable at $\Psi\sim\Psi_0$. When $\Psi$ decreases, the
first term decreases faster therefore well inside the jet one
can retain only the second term. Making use of Eqs. (\ref{Phi})
and (\ref{alpha}), one finds $\gamma=X$.
So in this case the Lorentz factor
of the flow is equal to the dimensionless cylindrical radius, which is the general property of the
equilibrium collimation.

When $\beta<1/4$, one substitutes the solution
(\ref{Ykappa=2b}) into Eq. (\ref{Lorentz-factor}) to yield the
relation
 \eqb
\frac{\eta^2\mu^2}{2\Omega^2\gamma^2}= \frac{\beta
C^2\sqrt{\alpha}}{Z^{2(1-k)}}
\int_{\widetilde{\Psi}}^{\Psi}\eta\mu\Phi^2 d\Psi+ \frac
1{\sqrt{\alpha}C^2Z^{2k}}
\int_{\widetilde{\Psi}}^{\Psi}\frac{\eta\mu}{\Phi^2}d\Psi.
 \label{gamma_beta<1/4}\eqe
At $\Psi\sim\Psi_0$,  the first term in the right-hand side
dominates the second one; this could be easily seen from Eq.
(\ref{Lorentz_lin}), which is the approximate form of Eq.
(\ref{Lorentz-factor}). Then one finds
 \eqb
\gamma=\frac{\eta\mu}{\Omega
C}\left(2\beta\sqrt{\alpha}\int_{\widetilde{\Psi}}^{\Psi}\eta\mu\Phi^2
d\Psi\right)^{-1/2}Z^{1-k};
 \eqe
which reproduces, in this specific case, the scaling
$\gamma\propto\sqrt{{\cal R}/r}$ common to the non-equilibrium
collimation. For $\mu$ given by the lineat function
(\ref{energy1}), this relation is reduced, with the aid of Eqs.
(\ref{Phi}) and (\ref{alpha_lin}) to
 \eqb
\gamma=\frac{3^{1/4}}{C}\sqrt{\frac{{\Psi_0}}{\beta\Psi}}Z^{1-k}.
 \label{gamma_noneq}\eqe
Note that in this case, the Lorentz factor increases, at a
fixed $Z$, towards the axis. The Lorentz factor increases until
at small enough $\Psi$, the second term in the right-hand side
of Eq. (\ref{gamma_beta<1/4}) becomes dominant, which means
that close enough to the axis, the jet is in pressure
equilibrium. In this region, one has
 \eqb
\gamma=\alpha^{1/4}C\sqrt{\frac{\Psi}{\widetilde{\Psi}}}Z^{k}=X;
 \eqe
as in any equilibrium flow. So at any fixed $Z$, the Lorentz
factor increases outwards from the axis while the flow is in
the pressure equilibrium and then decreases outwards.

The transition from the non-equilibrium to the equilibrium zone
occurs  at
 \eqb
\frac{\Psi}{\Psi_0}=\frac
1{\sqrt{\beta}C^2Z^{2k-1}};
 \eqe
when the two terms in the right-hand side of Eq. (\ref{gamma_beta<1/4}) become equal.
Transforming to the coordinate space with the aid of Eqs.
(\ref{jet_radius}) and (\ref{Ykappa=2b}), one sees that the transition occurs at the line
 \eqb
Z=\sqrt{\beta}X^{-2}.
 \label{parabola}\eqe
The Lorentz factor of the flow
increases as $\gamma=X$ while the flow remains within the line
(\ref{parabola}) whereas after the flow enters the non-equilibrium zone, the acceleration proceeds slower,
according to Eq. (\ref{gamma_noneq}).

\subsection{The case $\kappa>2$}
In this case, a solution to Eq.
(\ref{lin_app}) is presented as
  \eqb
w=\sqrt{Z}J_{\frac 1{\kappa-2}}\left(\frac{2\sqrt{\beta}}{\kappa-2}Z^{1-\kappa/2}\right).
 \label{solution_lin}\eqe
The large $Z$ asymptotics of this function corresponds to the small argument limit of the Bessel function,
 \eqb
J_{\nu}(x)=\frac 1{\Gamma(\nu+1)}\left(\frac x2\right)^{\nu}; 
 \eqe
where $\Gamma$ is the gamma-function. Then the auxiliary
function, $w$, goes to a constant \eqb w=\frac
1{\Gamma\left(\frac{\kappa-1}{\kappa-2}\right)}
\left(\frac{\sqrt{\beta}}{\kappa-2}\right)^{1/(\kappa-2)};
 \eqe
whereas the general solution (\ref{solution}) goes, at large $Z$, to a linear function
 \eqb
Y=\sqrt{C_1}\Gamma\left(\frac{\kappa-1}{\kappa-2}\right)
\left(\frac{\kappa-2}{\sqrt{\beta}}\right)^{1/(\kappa-2)}Z;
  \label{Ykappa>2_gen} \eqe
so that the flow becomes radial  at large distances.

One sees from Eq. (\ref{Ykappa>2_gen}) that the flow could be collimated, $Y\ll Z$, if $\kappa$ only
slightly exceeds 2 or/and $\beta$ is large. Note that in this case, the small argument limit of the Bessel
function, $x\ll 2$, which yields a linear asymptotics for $Y$, is achieved only at a very large $Z$.
For example, if $\kappa=2.5$, the above limit is achieved only at $Z\gg 16\beta^2$. In order to see
what happens at a smaller $Z$, let us assume that
 \eqb
 \frac{\sqrt{\beta}}{\kappa-2}\gg 1.
 \label{collimation_cond}\eqe
Then the argument of the Bessel function in Eq. (\ref{solution_lin}) is large at
 \eqb
Z\ll \left[\frac{2\sqrt{\beta}}{\kappa-2}\right]^{2/(\kappa-2)}.
 \label{argument}\eqe
In this case, one can use the large argument asymptotics of the Bessel function in Eq.
(\ref{solution_lin}), which leads to (cf. Eq. (\ref{Ykappa<2_gen}))
\eqb
w=\sqrt{\frac{\kappa-2}{\pi}}
\left(\frac{Z^{\kappa}}{\beta}\right)^{1/4}\cos S;\qquad
S=\frac{2\sqrt{\beta}}{\kappa-2}Z^{1-\kappa/2}-\frac{\kappa}{\kappa-2}\frac{\pi}4;
 \eqe
 \eqb
Y=\sqrt{\frac{\kappa-2}{\pi}}\left(\frac{Z^{\kappa}}{\beta}\right)^{1/4}\left[\frac
1{C_1}\cos^2S+C_1\left(C_2\cos S+ \frac{\pi}{2-\kappa}\sin S\right)^2\right]^{1/2}
 \label{Ykappa>2a}\eqe
So the solution obtained for the case $\kappa<2$ could be extended to $\kappa$ slightly above
2 but only in a limited range of $Z$. As in the case $\kappa<2$, the solution describes smooth expansion
 \eqb
Y=\left(\frac{Z^{\kappa}}{\beta}\right)^{1/4};
 \label{Ykappa>2}\eqe
only if the constants are specially chosen,
\eqb
C_1=(\kappa-2)/\pi;\quad C_2=0.
\label{constants}\eqe
Generally long wavelength oscillations are superimposed on the overall expansion.

One sees that if the condition (\ref{collimation_cond}) is
fulfilled, the flow is collimated according to Eqs.
(\ref{Ykappa>2a}) or (\ref{Ykappa>2}) in the region
(\ref{argument}); at a larger $Z$, the flow becomes radial
preserving the acquired collimation angle. In Fig. 1, solutions
to the governing equation are shown for $\kappa=2.5$,
$\beta=5$. These solutions are obtained by numerical
integration of Eq. (\ref{Y(Z)}), which is easier than numerical
evaluation of integrals in Eq. (\ref{solution}). The smoothly
expanded solution is shown by solid line whereas dashed line
represents a solution with oscillations in the region
(\ref{argument}). All the solutions go to a linear function at
large $Z$ where the flow is already well collimated. The final
collimation angle could be obtained from Eq.
(\ref{Ykappa>2_gen}) as $\Theta=Y/Z$. Choosing the constant
$C_1$ from Eq. (\ref{constants}) corresponding to the smoothly
expanded solution (\ref{Ykappa>2}), one finds the final
collimation angle as
 \eqb \Theta=\frac
1{\sqrt{\pi}}\Gamma\left(\frac{\kappa-1}{\kappa-2}\right)
\left(\frac{(\kappa-2)^{\kappa}}{\beta}\right)^{1/[2(\kappa-2)]}.
 \label{Theta}\eqe
One sees that the collimation angle rapidly increases with
increasing $\kappa$ and decreasing $\beta$;
$\Theta=0.01/\beta^{2.5}$ at $\kappa=2.2$, $\Theta=0.2/\beta$
at $\kappa=2.5$ and $\Theta=0.56/\sqrt{\beta}$ at $\kappa=3$.
This means that if the external pressure decreases faster than
${\cal P}\propto Z^{-3}$, a narrow jet could not be produced
unless the flow has already been collimated in the near zone,
$Z\sim 1$.

\begin{figure*}
\includegraphics[scale=0.8]{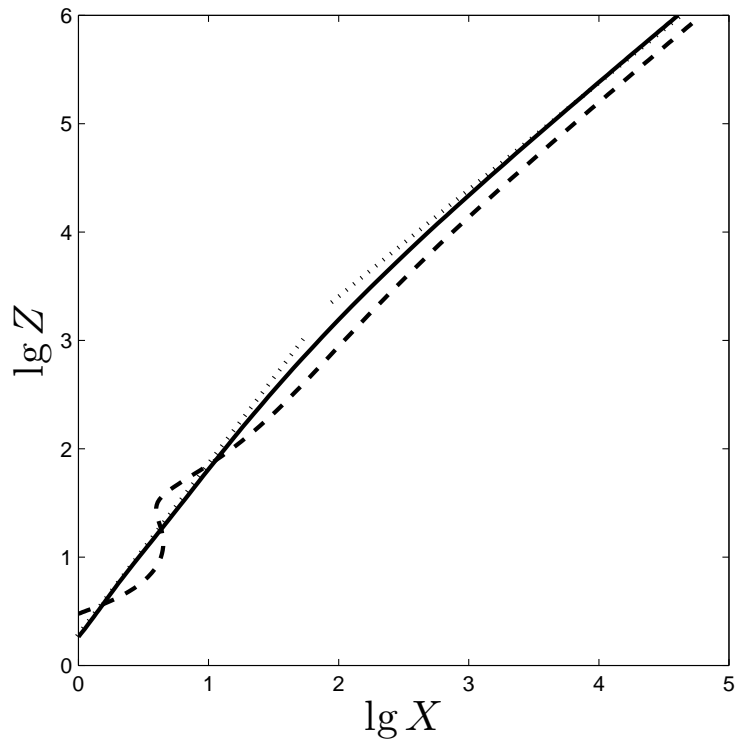}
\caption{The shape of the jet, $Y(Z)$, for $\kappa=2.5$;
$\beta=5$. The  solution without oscillations
is shown by solid line; dashed line shows a solution with
oscillations. Dotted lines show asymptotics, $X\propto
Z^{\kappa/4}=Z^{5/8}$ and $X\propto Z$ , correspondingly. }
\end{figure*}
\begin{figure*}
\includegraphics[scale=0.6]{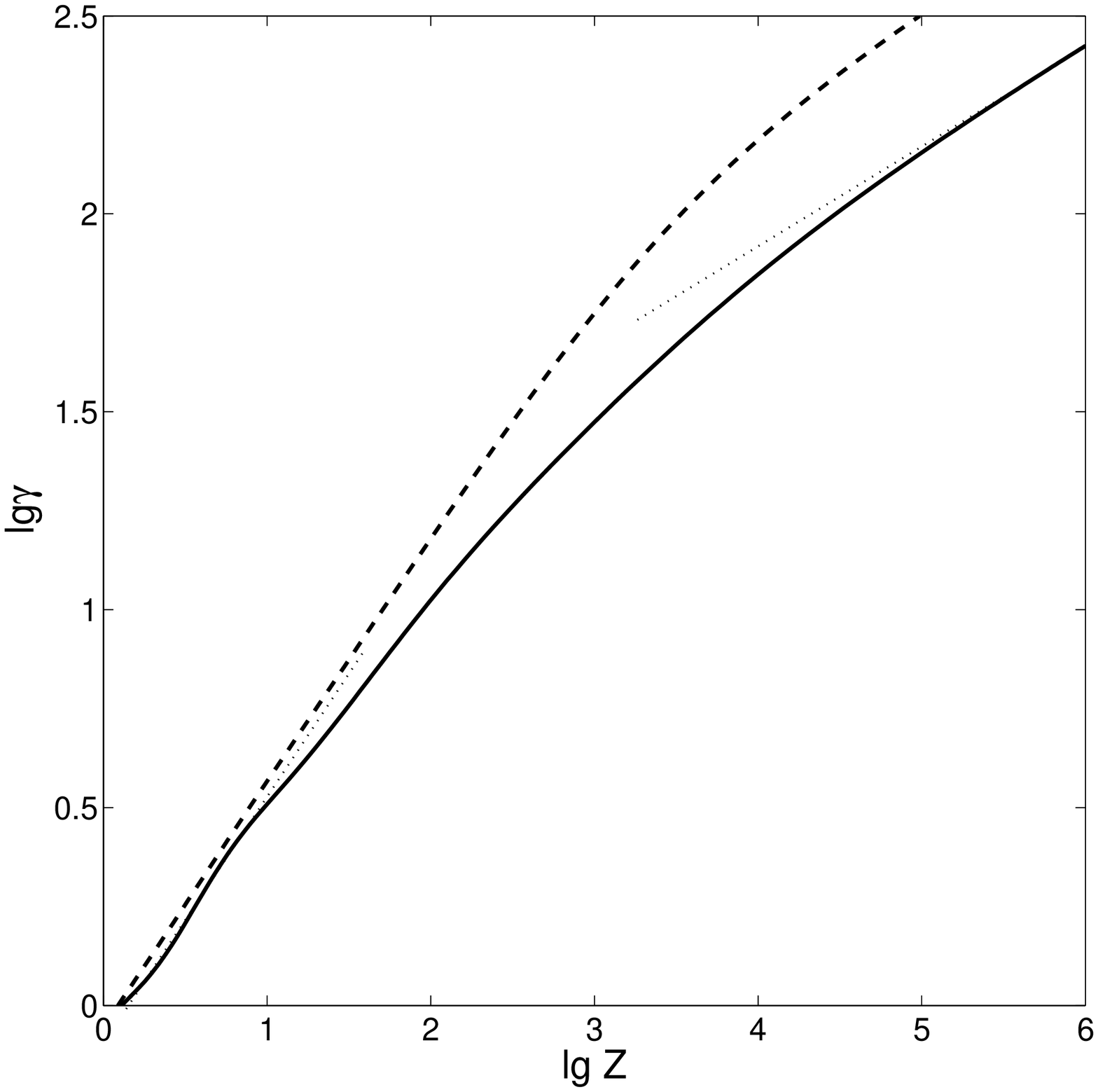}
\caption{Lorentz factor of the flow shown by solid line in Fig.
1 (expansion without oscillations). Solid line shows the
Lorentz factor of the flow at the boundary of the jet,
$\Psi=\Psi_0$; the dashed line is for the Lorentz factor at the
flux surface $\Psi=0.2\Psi_0$. Dotted lines show asymptotics,
$\gamma\propto Z^{\kappa/4}=Z^{5/8}$ and $\gamma\propto
Z^{(\kappa-2)/2}=Z^{1/4}$, correspondingly. }
\end{figure*}

Note that the solution (\ref{Ykappa>2}) corresponds to the
equilibrium collimation because it could be obtained by
neglecting the first term in the governing equation. In order
to figure out the collimation type of the radial flow
(\ref{Ykappa>2_gen}), one has to find the curvature of the
field surface. Expanding the solution to the governing equation
(\ref{Y(Z)}) to higher order terms in $1/Z$ (this could be
easier done by making expansion in the equation than by
expanding the general solution (\ref{solution}) and
(\ref{solution_lin})), one gets
 \eqb
Y=c_1Z+c_2+\left\{\begin{array}{ll}
\frac{\beta c_1}{(3-\kappa)(\kappa-2)}Z^{3-\kappa}; & 2<\kappa<4;
\\ \left(2c_1^3Z\right)^{-1}; & \kappa>4;\end{array}\right.
 \label{expanded_solution}\eqe
where $c_1$ and $c_2$ are constants, which could be expressed
via the constants $C_1$ and $C_2$ in the general solution
(\ref{solution}). Note that $c_1$ is in fact the final opening
angle of the jet, $c_1=\Theta$; below we will use the
expression (\ref{Theta}) for $c_1$. One sees from Eq.
(\ref{expanded_solution}) that at $\kappa>4$, the curvature is
independent of the parameters of the external pressure, $\beta$
and $\kappa$, which means that the flow is not confined by the
medium with such a sharply decreasing pressure. Below we do not
consider this case. At $\kappa<4$, the ratio of the first to
the second term in the governing equation (\ref{Y(Z)}) is
 \eqb
-Y^3\frac{d^2Y}{d^2Z}=\beta\Theta^4Z^{4-\kappa};
 \label{curv_vs_r}\eqe
which means that the flow is in the non-equilibrium regime at
 \eqb
Z>Z_1=\left(\beta\Theta^4\right)^{-1/(4-\kappa)}
 \eqe
For $\kappa=2.5$, the transition occurs at $Z_1=70\beta^2$.

The Lorentz factor of the flow is obtained, as before, from Eq.
(\ref{Lorentz-factor}) or, for the linear $\mu(\Psi)$, from Eq.
(\ref{Lorentz_lin}). At the condition (\ref{argument}), when
the flow is collimated according to Eq. (\ref{Ykappa>2}), the
second term in the right-hand side dominates the first one (the
equilibrium collimation) so that one comes again to Eq.
(\ref{gamma_kappa<2}), which is reduced, at $\Psi\ll\Psi_0$,
just to $\gamma=X$. At $Z>Z_1$, where the flow is
already nearly radial, the first term becomes dominating, which
yields
 \eqb
\gamma=\frac{\eta\mu}{\Omega
\Theta\alpha^{1/4}}\left(2\beta\int_{\widetilde{\Psi}}^{\Psi}\eta\mu\Phi^2
d\Psi\right)^{-1/2}Z^{(\kappa-2)/2}.
 \label{gamma_kappa>2gen}\eqe
This is in accord with the general non-equilibrium scaling $\gamma\propto\sqrt{{\cal R}/r}$.
For linear dependence of $\mu$ on $\Psi$, this relation is reduced, with the aid of Eqs. (\ref{Phi})
and (\ref{alpha_lin}), to
 \eqb
\gamma=\frac{3^{1/4}}{\Theta}\sqrt{\frac{\Psi_0}{\beta\Psi}}Z^{(\kappa-2)/2}.
 \label{gamma_kappa>2}\eqe
In Fig. 2, we show the Lorentz factor of the smoothly expanded
(without oscillations) flow at the same
parameters as in Fig. 1. One sees the transition from  a
relatively rapid acceleration in the equilibrium regime,
$\gamma\propto Z^{\kappa/4}=Z^{5/8}$ to the slow
non-equilibrium acceleration, $\gamma\propto
Z^{(\kappa-2)/2}=Z^{1/4}$.

An important point is that in the non-equilibrium regime, the
Lorentz factor increases with decreasing $\Psi$ so that the
flow is faster inside the jet than at the periphery. At small
enough $\Psi$, the second term in Eq. (\ref{Lorentz_lin}) could
become dominating, the transition occurring at
 \eqb
\frac{\Psi}{\Psi_0}=\left(\frac{Z_1}Z\right)^{(4-\kappa)/2}.
 \eqe
Recall that the non-equilibrium zone appears only at $Z>Z_1$.
In the coordinate space, the transition occurs at
 \eqb
X=\left(\frac{Z^{\kappa}}{\beta}\right)^{1/4}.
 \eqe
Inside the zone bounded by this surface, the flow is
accelerated as in the equilibrium case, $\gamma=X$.

One has to stress that at $Z>Z_1$, the Lorentz factor of the
flow is determined by the curvature of the flux surfaces
(non-equilibrium regime) even though the flux surfaces are
nearly conical so that the Lorentz factor depends on small
deviations from the conical shape. In this case, accuracy of
the governing equation could become insufficient in order to
find the Lorentz factor of the flow. The governing equation was
obtained by neglecting the kinetic energy term in the Bernoulli
equation (\ref{Bernoulli0}) therefore the shape of the flux
surfaces, $r(\Psi,z)$, is determined to within a factor of
$\gamma/\mu$. In the case of the equilibrium collimation,
$\gamma\propto r$, this accuracy is sufficient while the jet
remains Poytning dominated. In the case of non-equilibrium
collimation, the Lorentz factor goes as $\sqrt{{\cal R}/r}$ and if the
flow lines become nearly straight, the curvature could be
determined by the neglected terms of the order of $\gamma/\mu$.
In this case, the scalings (\ref{gamma_kappa>2gen}) and
(\ref{gamma_kappa>2}) cease to be valid when the flow is still
Poynting dominated. We address this issue in sect. 7.2.

\subsection{Comparison to previous works}
Recently magnetic acceleration of externally confined jets was
carefully studied, both numerically and analytically, by
\citet{tchekhovskoy08} and \citet{komissarov_etal08}.
\citet{tchekhovskoy08} used the force-free approximation
whereas \citet{komissarov_etal08} solved the full set of
relativistic MHD equations. For analytical estimates, \citet{tchekhovskoy08} assumed that
the shape of the flux surface is a power law and then found
from the transfield equation the appropriate Lorentz factor and
the external pressure. They obtained the scaling
(\ref{Ykappa<2}) and claimed that it is universal.
\citet{komissarov_etal08} based on the asymptotic transfield
equation obtained by \citet{vlahakis04}, which is equivalent to
our Eq. (\ref{asympt_transfield1}). Order of magnitude estimate
of terms in this equation led them to Eq. (\ref{Y(Z)}) for the
jet radius. Analyzing this equation, they revealed that the
scaling (\ref{Ykappa<2}) is valid only for $\kappa<2$. For
$\kappa=2$, they obtained the scalings (\ref{Ykappa=2a}) and
(\ref{Ykappa=2b}) for $\beta>1/4$ and $\beta<1/4$,
correspondingly. For $\kappa>2$, they obtained the radial
asymptotics.

Our approach generalizes these findings permitting the
asymptotically exact solutions describing the full structure of
the get. Going beyond the simplest power law scalings also
permit us to find important new qualitative features of the
flow. In particular, we found that in the case of the
equilibrium collimation, $\kappa<2$, oscillations could be
superimposed on the general expansion of the jet. For
$\kappa>2$, we see the transition between the equilibrium and
non-equilibrium regimes, which could not be described by a
power law scaling.  Namely, if $\kappa$ only slightly exceeds
2, the flow is collimated according to the equilibrium law
(\ref{Ykappa<2}) but only till some limiting distance beyond
which the flow becomes radial preserving the acquired
collimation angle. The larger $(\kappa-2)$, the earlier (at a
smaller distance) the flow becomes radial so that at $\kappa>3$
the flow is practically
radial from the very origin. 

\citet{tchekhovskoy08} numerically simulated jets with
different profiles of the external pressure.
They found excellent agreement with the scaling
(\ref{Ykappa<2}) at $\kappa=2$. For $\kappa=2.5$ they reported noticeable
deviations from this scaling at large distances; we suppose that they
observed the transition to the radial flow. At last for
$\kappa=2.8$, they found a wide conical jet. These results agree
with our conclusions.

\section{Terminal Lorentz factor and collimation angle.}

It was shown in the previous section that if the Poynting
dominated outflow is confined by the external pressure, the
flow is collimated and the plasma is accelerated so that
eventually the kinetic energy of the plasma could not be
neglected any more. An important point is that the closer to
the axis, the earlier (at a smaller $z$) this happens. Note that at
$\Psi\lesssim\widetilde{\Psi}$, the Poynting flux is relatively
not large from the very beginning. In this section, we address saturation
of the acceleration  and estimate the terminal Lorentz
factor and collimation angle. For
the estimates, we assume that the energy integral is described
by the linear function (\ref{energy1}); then the parameters of
the flow are given by Eqs. (\ref{Phi}), (\ref{alpha_lin}) and
(\ref{beta_lin}). It is also convenient to introduce the
maximal achievable Lorentz factor,
 \eqb
 \gamma_{\rm max}=\mu(\Psi_0)\approx\gamma_{\rm in}\Psi_0/{\widetilde
\Psi};
 \eqe
which is just Michel's magnetization parameter
\citep{michel69}.

The results of the previous section have been obtained in the
limit $\gamma\ll\mu$, which means that the shape of the flux
surfaces has been found with the accuracy of $\gamma/\mu$. We
showed that the flow is accelerated as $\gamma\sim X$ in the
case of equilibrium collimation, i.e. under the condition
(\ref{equilibrium_cond}), and as $\gamma\propto\sqrt{{\cal R}/r}$ in the
opposite case. A small error in the shape of the
flux surface yields the error of the same order in
the equilibrium scaling $\gamma\propto X$ therefore this scaling could be
safely extrapolated up to $\gamma\sim\mu\sim\gamma_{\rm in}\Psi/\Psi_0$,
i.e. until the flow ceases to be Poynting dominated. In the case of
non-equilibrium collimation, one has to analyze how corrections
of the order of $\gamma/\mu$ could alter the curvature, $1/{\cal
R}=-d^2r/dz^2$. Small corrections to the
shape of the flux surface could significantly modify the
curvature if the flux surfaces are close to cones, i.e. if
$r(Z)$ is close to a linear function. This indeed happens when
the confining pressure decreases faster than $z^{-2}$.
Therefore we consider separately the cases $\kappa\le 2$ and
$\kappa>2$.

\subsection{The case $\kappa\le 2$; transition to $\sigma\sim 1$.}

In the case $\kappa<2$ the Lorentz factor of the flow increases
according to the equilibrium law $\gamma=X$. Corrections of the
order of $\gamma/\mu\ll 1$ could not affect significantly this
scaling therefore the flow is accelerated up to
$\gamma\sim\mu\sim\gamma_{\rm max}\Psi/\Psi_0$; this occurs at
$X\sim\gamma_{\rm max}\Psi/\Psi_0$. Making use of Eq.
(\ref{r_linear}) and (\ref{Ykappa<2}), one finds the
corresponding distance as
 \eqb
Z=\left[\frac{\beta}3\gamma_{\rm max}^4\left(\frac{\Psi}{\Psi_0}\right)^2\right]^{1/\kappa}.
 \eqe
Reverting this expression, one finds the boundary of the moderately magnetized, $\sigma\sim 1$, core as
 \eqb
\frac{\Psi_{\rm core}}{\Psi_0}=\sqrt{\frac 3{\beta}}\frac{Z^{\kappa}}{\gamma^2_{\rm max}}.
 \eqe
In the coordinate space, the boundary of the moderately magnetized
core is found as
 \eqb
X_{\rm core}=\sqrt{\frac{3}{\beta}}\frac{Z^{\kappa/2}}{\gamma_{\rm max}}.
 \label{Poynting_boundary}\eqe

One sees that the closer to the axis, the smaller the distance
where the acceleration saturates so that a mildly
magnetized, $\sigma\sim 1$, core expands with the distance
occupying a progressively larger fraction of the jet body until
the $\sigma\sim 1$ state is achieved across the whole  jet. This
happens at the distance
 \eqb
Z_{\sigma}= \left(\frac{\beta}3\gamma_{\rm max}^4\right)^{1/\kappa}
 \eqe
from the origin. At this distance, the collimation angle, $\Theta=dY/dZ$, is
 \eqb
\Theta=\frac{3^{(4-\kappa)/4\kappa}\kappa}{4\beta^{1/4}}\gamma_{\rm
max}^{-(4-\kappa)/\kappa}.
 \eqe
When the flow ceases to be Poynting dominated, the collimation angle decreases further. It will be shown elsewhere (Lyubarsky, in preparation) that if $\kappa<2$, the flux surfaces become cylindrical at the infinity and the Poynting flux is totally transferred to the kinetic energy.

If the density decreases not as a power law (but slower than $z^{-2}$), one can write general estimates making use of Eq. (\ref{Ygeneral}) and (\ref{gamma_p}). Specifically one finds that the pressure should decrease
at least by a factor of $\beta\gamma_{\rm max}^4$ in order for the Poynting flux to be converted into the kinetic energy of the flow. If the environment at large distances from the compact object has the finite pressure $p_{\rm amb}>p_0/(\beta\gamma_{\rm max}^4)\sim B_0^2/(6\pi\gamma_{\rm max}^4)$, the flow becomes cylindrical and the Lorentz factor is saturated at the value
 \eqb
 \gamma_t=\left(\frac{B_0^2}{6\pi p_{\rm amb}}\right)^{1/4}.
\eqe

The above estimates are illustrated by a sketch in Fig. 3. The curve 1 shows the distribution of the Lorentz factor across the jet at a not very large distance from the origin where the flow is still Poynting dominated everywhere with except of the region $\Psi\lesssim\widetilde {\Psi}$. Further out of the origin, the $\sigma\sim 1$ core expands within the jet. The curve 2 shows the Lorentz factor at some intermediate
distance where the core is already developed but the main body of the jet is still Poynting dominated.
The curve 3 sketches the distribution of the Lorentz factor at the distance $Z\sim Z_{\sigma}$ when the whole jet ceases to be Poynting dominated.

\begin{figure*}
\includegraphics[scale=0.6]{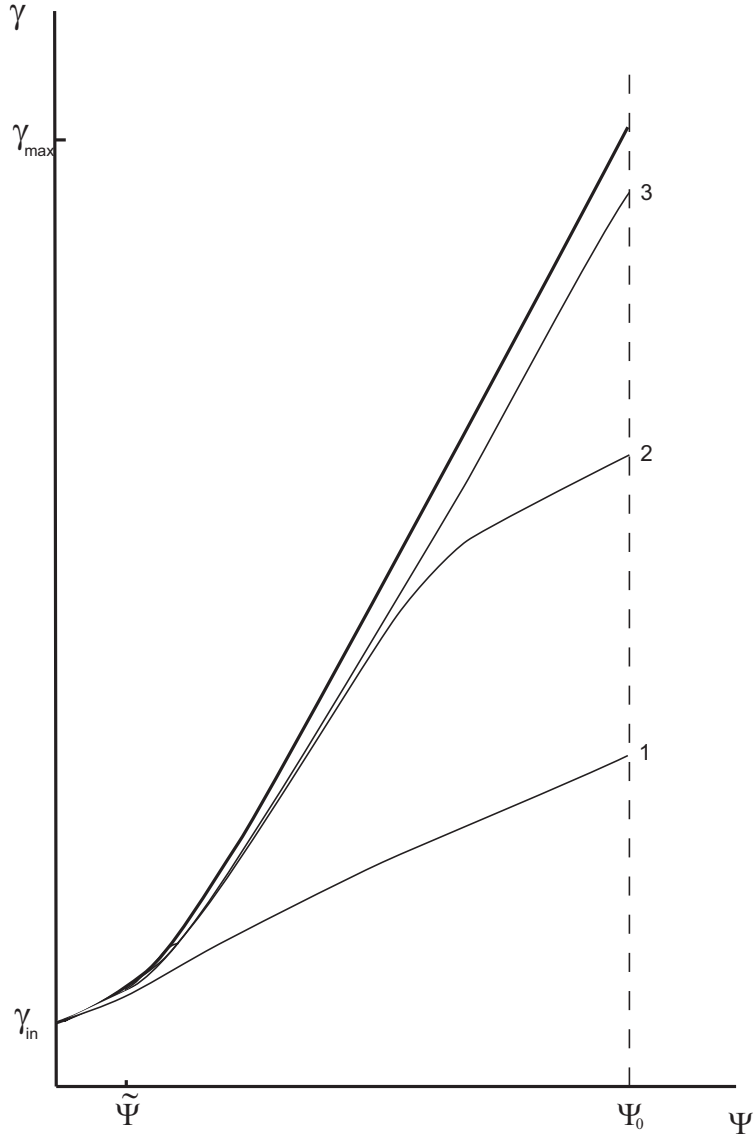}
\caption{Distribution of the Lorentz factor across the jet in the case $\kappa<2$; not to scale.
Thick line shows the distribution of the total energy, $\mu(\Psi)$. Each thin line
shows the distribution of the Lorentz factor at some distance from the origin; they are labeled
according to the distance, i.e. the curve 1 is the closest to the origin, the curve 3 is the farthest.}
\end{figure*}

The above estimates could also be applied to the case
$\kappa=2$, $\beta>1/4$ when the flow follows scalings
obtained for $\kappa<2$; one has just substitute $\kappa$ by 2 and
$\beta$ by $\beta-1/4$. For example, the boundary of the moderately magnetized core is now presented
as
 \eqb
X_{\rm core}=\sqrt{\frac 3{\beta-1/4}}\frac Z{\gamma_{\rm max}};
 \label{core_boundary}\eqe
whereas the distance where the jet ceases to be Poynting dominated is
 \eqb
Z_{\sigma}=\sqrt{\frac{\beta-1/4}3}\gamma^2_{\rm max}.
 \eqe

On the contrary, if $\kappa=2$ and
$\beta<1/4$, the flow exhibits quite different behavior because
it is collimated in the non-equilibrium regime. In such a flow,
the Lorentz factor grows $\propto X$ only if and while the flow
line remains close enough to the axis where the flow is in the
equilibrium. After the flow line crosses the boundary between
the equilibrium and non-equilibrium zone given by Eq.
(\ref{parabola}), the acceleration proceeds slower, the Lorentz
factor being given by Eq. (\ref{gamma_noneq}). In the Poynting
dominated domain, the Lorentz factor is not monotonic across
the jet; at a fixed $Z$, it increases with the radius within
the equilibrium zone and decreases outwards in the
non-equilibrium zone. Therefore at any $Z$, the Lorentz factor reaches
the maximum value somewhere within the jet.  Comparing the Lorentz
factor with $\mu(\Psi)$, one finds that at a distance $Z$ from the
origin, the flow remains Poynting dominated only outside the boundary
 \eqb
\frac{\Psi_{\rm core}}{\Psi_0}=\left\{\begin{array}{ll}
\sqrt{3}C^2\gamma_{\rm max}^{-2} Z^{2k}; & Z<\left[\gamma_{\rm max}/(3^{1/4}\beta^{1/4}C^2)\right]^{2/(4k-1)};
\\ 3^{1/6}\left(\beta C^2\gamma_{\rm max}^2\right)^{-1/3}Z^{(2/3)(1-k)}; &
Z>\left[\gamma_{\rm max}/(3^{1/4}\beta^{1/4}C^2)\right]^{2/(4k-1)}.\end{array}\right.
 \eqe
or, in the coordinates,
 \eqb
X>\left\{\begin{array}{ll}
\sqrt{3}C^2\gamma_{\rm max}^{-1} Z^{2k}; & Z<\left[\gamma_{\rm max}/(3^{1/4}\beta^{1/4}C^2)\right]^{2/(4k-1)};
\\ \left(3C^2\beta^{-1/2}\gamma_{\rm max}^{-1}Z^{2k+1}\right)^{1/3}; &
Z>\left[\gamma_{\rm max}/(3^{1/4}\beta^{1/4}C^2)\right]^{2/(4k-1)}.\end{array}\right.
 \eqe
One sees that the acceleration in the central part of the jet should be rapidly
saturated because the Lorentz factor achieves the maximal value $\gamma\sim\mu(\Psi)$.
Slow acceleration in the outer, non-equilibrium part of the jet continues until the Lorentz
factor approaches $\mu(\Psi)$. The closer the flow line to the axis, the earlier
(at a smaller $Z$) this happens so that in this case also a moderately magnetized
core expands within the body of the jet. At $Z>\left[\gamma_{\rm max}/(3^{1/4}\beta^{1/4}C^2)\right]^{2/(4k-1)}$, a maximal Lorentz factor is
achieved at the boundary of the core because in the non-equilibrium jet, the acceleration
is faster at the flow lines closer to the axis. The whole jet ceases to be Poynting dominated at the distance
 \eqb
Z_{\rm conv}=\left(3^{-1/4}\sqrt{\beta}C\gamma_{\rm max}\right)^{1/(1-k)}.
 \eqe
The corresponding collimation angle is
 \eqb
\Theta=\frac{3^{1/4}k}{\sqrt{\beta}\gamma_{\rm max}}.
 \eqe

 \begin{figure*}
\includegraphics[scale=0.6]{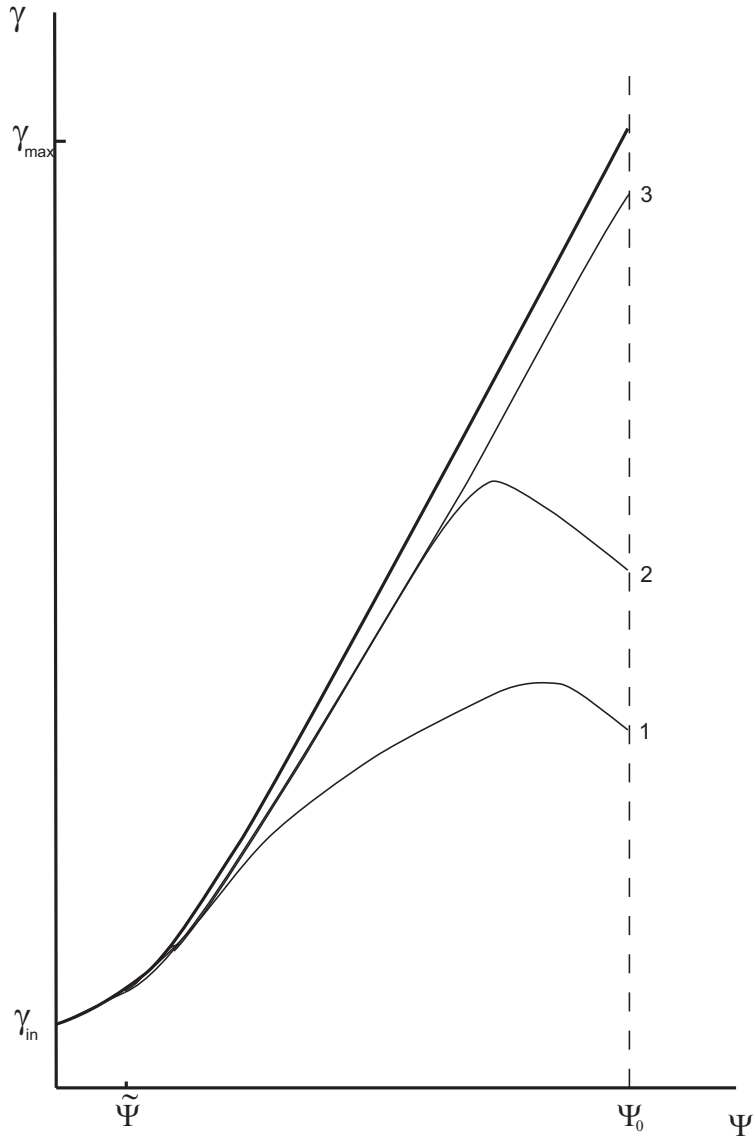}
\caption{The same as in Fig. 3 but in the case $\kappa=2$, $\beta<1/4$.}
\end{figure*}

The distribution of the Lorentz factor across the jet in the case $\kappa=2$; $\beta<1/4$
is sketched in Fig. 4. The curve 1 shows the Lorentz factor not far from the origin. An internal part
of the jet is collimated in the equilibrium regime and the Lorentz factor increases outwards from the axis.
In the main body of the jet, the collimation is non-equilibrium and the Lorentz factor decreases
outwards so that the Lorentz factor is a maximum at the boundary between the equilibrium and non-equilibrium zones. As the jet propagates, the flow in the internal parts reaches the $\sigma\sim 1$ state and stops
accelerating. The curve 2 shows the distribution of the Lorentz factor at the distance $Z>\left[\gamma_{\rm max}/(3^{1/4}\beta^{1/4}C^2)\right]^{2/(4k-1)}$
where the flow outside the core is non-equilibrium so that the maximal Lorentz factor is achieved
at the boundary of the core. The curve 3 shows the distribution of the Lorentz factor at $Z\sim Z_{\rm conv}$ when the whole jet ceases to be Poynting dominated.

\subsection{The case $\kappa>2$; transition to logarithmic acceleration.}

At $\kappa>2$, the flow is collimated only if the condition
(\ref{collimation_cond}) is fulfilled. In this case, the flow
becomes conical still being Poynting dominated; the
final collimation angle is given by Eq. (\ref{Theta}). In
the conical part of the jet, the Lorentz factor grows according
to Eq. (\ref{gamma_kappa>2}), which corresponds to the general
non-equilibrium scaling $\gamma\propto\sqrt{{\cal R}/r}$. An
important point is that the curvature of the flow lines is
determined by small deviations from the straight line, see Eq.
(\ref{expanded_solution}), therefore Eq.(\ref{gamma_kappa>2})
is valid only if the shape of the flow lines could be found from the governing equation
with the necessary accuracy. Let us consider more carefully the jet with nearly
straight flux surfaces.

In the previous sections, we found the jet structure neglecting
$\gamma$ as compared with $\mu$  in the Bernoulli equation
(\ref{Bernoulli0}). Then the shape of the flux surfaces is
presented as (see Eqs. (\ref{jet_radius}) and (\ref{r_linear}))
$X(\Psi,Z)=\alpha^{1/4}\Phi(\Psi)
Y(Z)=3^{1/4}\sqrt{\Psi/\Psi_0}Y(Z)$, where $Y(Z)$ satisfies the
governing equation. We can find limits of applicability of Eq.
(\ref{gamma_kappa>2}) for the Lorentz factor of the flow by substituting
this equation into the Bernoulli
equation (\ref{Bernoulli0}) and finding the corresponding
corrections to the shape of the flux surfaces. Eq.
(\ref{gamma_kappa>2}) is valid while the curvature due to this
corrections remains small as compared with the curvature
obtained from the solution (\ref{expanded_solution}) of the
governing equation.

Let us present the shape of the flux surfaces as (cf. Eq. (\ref{jet_radius}))
 \eqb
X=\alpha^{1/4}\Phi Y(1+\delta);
 \label{X_corr}\eqe
where $\delta(\Psi,Z)\ll 1$ describes corrections to the shape of the flux
surfaces due to a non-zero $\gamma/\mu$. Substituting this into
(\ref{Bernoulli0}) and linearizing with respect to small $\delta$ and
$\gamma/\mu$, one gets
 \eqb
\frac{\partial\delta}{\partial\Psi}=\frac{\Omega^2\gamma}{\eta\mu^2}.
 \eqe
Assuming for simplicity that the energy integral is
described by the linear function (\ref{energy1}), one
writes in the dimensionless form
 \eqb
\frac{\partial\delta}{\partial S}=\frac{\gamma}{2\gamma_{\rm
max}S^2};
  \label{corr_Bern}\eqe
where
 \eqb
S=\Psi/\Psi_0.
 \eqe
With $\gamma$ from Eq. (\ref{gamma_kappa>2}), one finds
 \eqb
\delta=-\frac{3^{1/4}Z^{(\kappa-2)/2}}{2\Theta\sqrt{\beta}\gamma_{\rm
max}S^{3/2}}.
 \eqe
Substituting this expression into Eq. (\ref{X_corr}) and differentiating twice
with respect to $Z$, one finds the curvature of the flux surface as
 \eqb
\frac{d^2X}{dZ^2}=3^{1/4}\sqrt{\frac{\Psi}{\Psi_0}}\left(\frac{d^2Y}{dZ^2}-
\frac{ 3^{1/4}\kappa(\kappa-1)}{4\sqrt{\beta}\gamma_{\rm
max}S^{3/2}Z^{2-\kappa/2}}\right).
 \eqe
Here we take into account that $Y\approx\Theta Z$. The Lorentz
factor of the flow could be determined from the solution to the
governing equation only if the second term in brackets is small
as compared with the first one. Finding $d^2Y/dZ^2$ from  Eq.
(\ref{expanded_solution}), one sees that this is the case only
at distances smaller than
 \eqb
Z_t(\Psi)=
\left[\frac{4\gamma_{\rm
max}\Theta}{3^{1/4}\kappa(\kappa-1)}
\right]^{2/[3(\kappa-2)]}\left(\frac{\beta\Psi}{\Psi_0}\right)^{1/(\kappa-2)}.
 \label{Zt}\eqe
At this distance, the flow acquires the Lorentz factor
 \eqb
\gamma_t=\left(\frac{4\sqrt{3}}{\kappa(\kappa-1)}\frac{\gamma_{max}}{\Theta^2}\right)^{1/3}.
 \eqe
Note that this Lorentz factor is the same for all flux
surfaces whereas $Z_t$ increases towards the periphery of the
jet. This is because in the non-equilibrium regime, the
acceleration rate decreases outwards from the axis.

Let us now find the Lorentz factor of the flow at $Z>Z_t$. With this
purpose, one has to solve the transfield and
the Bernoulli equations without neglecting $\gamma$ in the Bernoulli equation.
In the case of interest, the collimation
is non-equilibrium therefore one can take the transfield
equation in the form (\ref{nonequilibrium}). As the flow lines are nearly straight,
we can look for the
solution in the form (\ref{X_corr}) with $Y(Z)=\Theta Z$. We
again assume for simplicity that the energy integral is a
linear function (\ref{energy1}); then $\Phi$ and $\alpha$ are
given by Eqs. (\ref{Phi}) and (\ref{alpha_lin}),
correspondingly. Now the transfield equation is written in the
limit $\delta\ll 1$, $\gamma/\mu\ll 1$ as
 \eqb
-\sqrt{3}\Theta^2Z\left(2\frac{\partial\delta}{\partial Z}+
\frac{\partial^2\delta}{\partial Z^2}\right) =\frac 2{S\gamma}\frac{\partial}{\partial S}\frac S{\gamma}.
 \label{transfield_corr}\eqe
Linearization of the Bernoulli equation in small $\delta$ and
$\gamma/\mu$ yields Eq. (\ref{corr_Bern}).

Eliminating $\delta$ from
these two equations (by differentiating Eq.
(\ref{transfield_corr}) with respect to $S$ and substituting
Eq. (\ref{corr_Bern})), one gets a single equation for
$\gamma$:
 \eqb
\frac{\sqrt{3}\Theta^2}{4\gamma_{max}}\gamma^2
\left(2Z\frac{\partial\gamma}{\partial Z}+Z^2\frac{\partial^2\gamma}{\partial Z^2}\right)=
1+2\frac S{\gamma}\frac{\partial\gamma}{\partial S}+
S^2\gamma^2\frac{\partial}{\partial S}
\left(\frac 1{\gamma^3}\frac{\partial\gamma}{\partial S}\right).
 \label{gamma}\eqe
As an initial condition, one can take $\gamma=\gamma_t$ at
$Z=Z_t$ (the above estimates give in fact $\gamma\sim\gamma_t$ at
$Z\sim Z_t$). Note that $\gamma_t$ is independent of
$S$. Assuming that beyond $Z_t$, the solution is also independent of $S$, one
finds with the logarithmic accuracy, i.e. in the
limit $\ln Z\gg 1$,
 \eqb
\gamma=\left[\frac{2\sqrt{3}\gamma_{max}}{\Theta^2}(\ln
CZ)\right]^{1/3}.
 \eqe
One sees that with the constant $C=1/Z_t$, this function
goes to $\gamma\sim\gamma_t$ at $Z\sim Z_t$ and still satisfies Eq. (\ref{gamma})
with the logarithmic accuracy (because it only  logarithmically depends on $S$, via $Z_t$).
So the final solution at $Z\gg Z_t$ (in fact at $\ln Z/Z_t\gg 1$) is written as
 \eqb
\gamma=\left(\frac{2\sqrt{3}\gamma_{max}}{\Theta^2}\ln
\frac{Z}{Z_t}\right)^{1/3}.
 \eqe
According to this solution, the flow in
fact stops accelerating beyond the distance $Z_t$ so that one
can use $\gamma_t$ as an estimate for the terminal Lorentz
factor. This conclusion matches with the well-known result
\citep{tomimatsu94,beskin_etal98} that the radial, non-confined
wind is accelerated only till
$\gamma\sim\gamma_{\rm max}^{1/3}$ and then the Lorentz factor
grows only as $(\ln R)^{1/3}$.

One has to stress that according to the boundary condition
(\ref{boundary}), the flow at the boundary is accelerated till
$\gamma_{\rm max}$ provided the external pressure falls to
zero. Therefore close enough to the boundary of the flow, the
Poynting flux is efficiently converted into the kinetic energy.
Specifically for the power-law pressure profile
(\ref{pressure}), the boundary condition (\ref{boundary})
yields in the limit $\gamma\ll \gamma_{\rm max}$
(\ref{energy1})
 \eqb
\gamma(\Psi_0,Z)=\frac{3^{1/4}}{\beta^{1/2}\Theta}Z^{(\kappa-2)/2};
 \label{accel_boundary}\eqe
which recovers the scaling (\ref{gamma_kappa>2}).  On sees that
at the boundary of the flow, the scaling (\ref{gamma_kappa>2})
remains valid until $\gamma\sim\gamma_{\rm max}$ even though in
the main body of the jet, the acceleration is saturated at
$\gamma\sim\gamma_t$. In order to find the width of the
boundary region where the acceleration proceeds beyond
$\gamma_t$, let us substitute Eq. (\ref{accel_boundary}) into
Eq. (\ref{gamma}) and estimate $\partial^2\gamma/\partial
S^2\sim \gamma/(\Delta S)^2$ necessary to satisfy the equation.
This yields
 \eqb
\Delta S=\frac{\Delta\Psi}{\Psi_0}\sim
\left(\frac{Z_t(\Psi_0)}Z\right)^{(3/4)(\kappa-2)}.
 \eqe
One sees that after the flow reaches $\gamma\sim\gamma_t$ at
$Z\sim Z_t$, the acceleration proceeds further only in a narrow
region close to the boundary.

 \begin{figure*}
\includegraphics[scale=0.6]{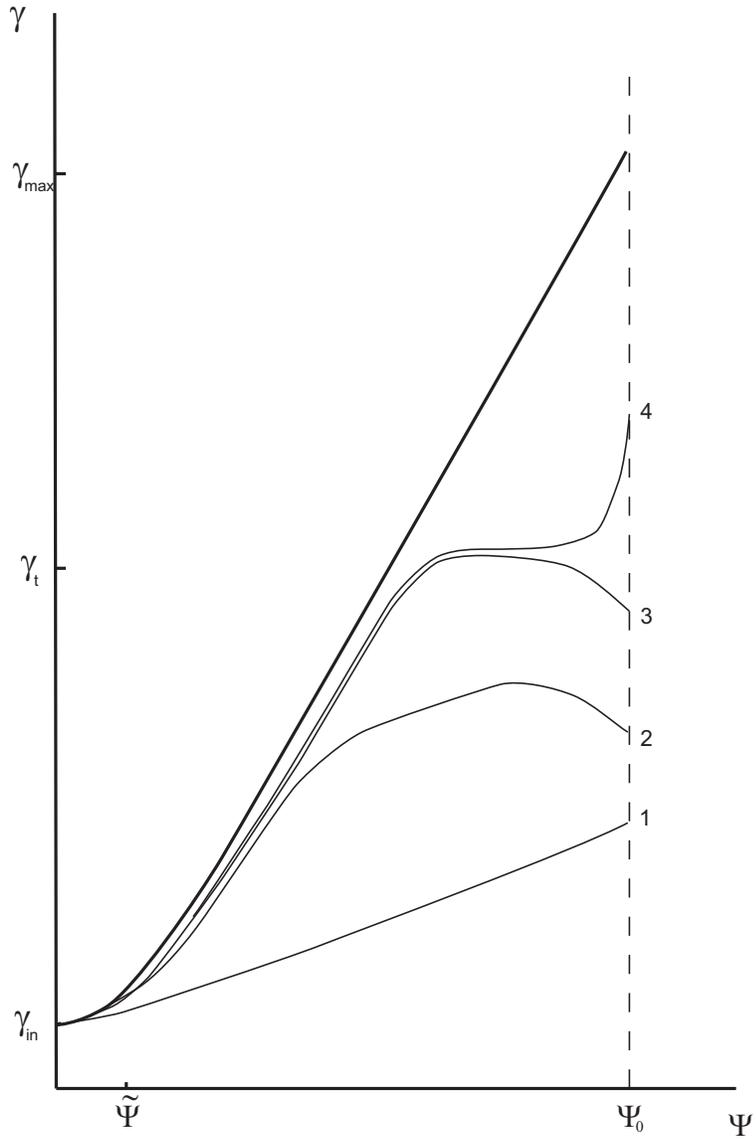}
\caption{The same as in Fig. 3 but in the case $\kappa>2$.}
\end{figure*}

The results of this subsection are illustrated in Fig. 5. The curve 1 shows the distribution
of the Lorentz factor at some distance $Z<Z_1$ where the jet is collimated in the equilibrium regime.
The curve 2 corresponds to a distance $Z>Z_1$ where the main body of the jet is in the non-equilibrium regime
so that $\gamma$ has a maximum inside the jet, at the boundary between the equilibrium and the non-equilibrium zones. The curve 3 shows the distribution of the Lorentz factor at $Z=Z_t(\Psi)$ for some $\Psi<\Psi_0$.
At this distance, the acceleration is saturated in the internal part of the jet. At last the curve 4
shows the Lorentz factor at $Z>Z_t(\Psi_0)$ where the acceleration is saturated in the main body of
the jet and proceeds further only in a narrow boundary layer.

Note also that the above estimates assume that the flow remains Poynting
dominated, i.e., that $\gamma_t<\gamma_{\rm max}\Psi/\Psi_0$.
In the opposite case, the acceleration is saturated at
$\gamma\sim\gamma_{\rm max}\Psi/\Psi_0$. In particular, if
$\gamma_t>\gamma_{\rm max}$, the total Poynting flux is
efficiently converted into the kinetic energy. The last
condition could also be written as $\gamma_{\rm max}\Theta<1$,
which means that the flow remains causally
connected in the sense that a signal sent in the transverse direction (in the proper plasma frame)
could cross the jet for the proper time $z/\gamma$. Recall that the same condition
is satisfied when $\kappa\le 2$ therefore in any case, the whole flow is
accelerated till $\sigma\sim 1$ only if it remains causally connected.
If the acceleration is saturated at $\gamma\sim\gamma_t<\gamma_{\rm max}$, the flow
is causally disconnected, $\gamma_t\Theta\sim (\gamma_{\rm max}\Theta)^{1/3}>1$.
By this reason, in particular, the flow do not "feel" the boundary any more so that
the acceleration stops everywhere except of a narrow boundary region.
The linkage between acceleration and causal connection of the flow was also noted by \citet{tchekhovskoy09}. The loss of causal connection also implies the global MHD stability of such jets
because global instabilities (e.g., the kink instability) has no time to develop.

Note also that $\Theta$ is determined only by the
external pressure profile whereas $\gamma_{\rm max}$ is determined by
the parameters of the outflow so that these two quantities are
independent and any relation between them is possible. However,
$\gamma_t\Theta$ could hardly ever be very large in real systems because
this quantity depends on the parameters in the power 1/3. On the other hand,
such a flow could be accelerated further when and if the Poynting flux
is dissipated \citep{thompson94,lyubarsky_kirk01,drenkhahn02,drenkhahn_spruit02,kirk_olaf03}.

One should also note in this connection that according to a
widely accepted view, the observed achromatic breaks in the GRB
afterglow light curves occur when $\Theta\gamma$ becomes
approximately unity. Since the afterglow is attributed to the
decelerating jet, this implies that $\gamma\Theta$ was
larger than unity in the prompt phase. One sees
that the required property could be achieved in the MHD
scenario if the confining pressure decreases with the distance
something faster than $z^{-2}$.

\section{The core of the jet}

It was shown in the previous section that a moderately
magnetized core occurs near the axis of a Poynting dominated
flow so that the solutions obtained in the section 6 could not
be continued to the axis. In this section, we find the structure of such a
core smoothly matched with the structure of the Poynting dominated flow in the main body of the jet.

The flow near the axis is described by Eq.
(\ref{axis_transfield}), which should be complemented by the
Bernoulli equation in the form (\ref{Bernoulli0}):
 \begin{eqnarray}
\mu\left(1+\frac{\gamma_{\rm
in}^2}{\gamma^2}\right)-2\gamma_{\rm in}= (X^2+\gamma_{\rm
in}^2)\frac{\mu-\gamma}{\Omega^2\gamma}
\frac{\partial}{\partial\Psi}\frac{\eta\mu}{\gamma};
 \label{core_tr}\\
\eta (\mu-\gamma)\frac{\partial X}{\partial\Psi}=\Omega^2 X.
 \label{core_Bern}\end{eqnarray}
This set of the first order differential equations for
$X(\Psi)$ and $\gamma(\Psi)$ should be solved at the condition
$X(0)=0$ and matched, at large $\Psi$, to the solution in the main body of the flow. For
example, if the main body of the flow is Poynting dominated, the solution to Eqs.
(\ref{core_tr}) and (\ref{core_Bern}) should be matched with the solutions obtained
in Section 6. The dependence on $Z$ enters only via this matching.

Close to the axis, the energy integral has the form of a
linear function (\ref{energy1}). Introducing the variables
 \eqb
s=1+\frac{\Psi}{\widetilde{\Psi}};\quad \xi=\frac X{\gamma_{\rm
in}};\quad \Gamma=\frac{\gamma}{\gamma_{\rm in}};
  \eqe
one reduces Eqs. (\ref{core_tr}) and (\ref{core_Bern}) to
dimensionless form
  \begin{eqnarray}
\frac s{\Gamma}\frac{d\Gamma}{ds} & = & 1-\frac{s+\Gamma^2(s-2)}{2(1+\xi^2)(s-\Gamma)};\label{core_tr1}\\
\frac{d\xi}{ds} & = & \frac{\xi}{2(s-\Gamma)}.\label{core_Bern1}
  \end{eqnarray}
Before presenting the numerical solution to these equations,
let us investigate them qualitatively.

Near the axis, $\xi\ll 1$, the solution is
 \eqb
s=1+C\xi^2;\qquad \Gamma=1+\frac 12 C\xi^4;
 \label{init}\eqe
where $C$ is a constant. This means, in particular, that the
poloidal magnetic field is homogeneous at
$\Psi\ll\widetilde{\Psi}$.

Far from the axis, $s\gg 1$, $\xi\gg 1$, the flow is
accelerated, $\Gamma\gg 1$, so that Eq. (\ref{core_tr1}) is
reduced to
 \eqb
\frac
s{\Gamma}\frac{d\Gamma}{ds}=1-\frac{s\Gamma^2}{2\xi^2(s-\Gamma)}.
 \label{core_tr-asympt}\eqe
The set of equations (\ref{core_Bern1}) and
(\ref{core_tr-asympt}) is invariant with respect to the
transformation $s\to\lambda s$; $\xi\to\lambda\xi$;
$\Gamma\to\lambda\Gamma$ so that
the equations could be integrated. Namely, introducing the
variables
 \eqb
u=\frac{\Gamma}{\xi};\qquad \sigma=\frac{s}{\Gamma}-1;
 \label{newvar}\eqe
(note that $\sigma$ thus defined is indeed the ratio of the
Poynting to the kinetic energy fluxes) yields the set of
equations
  \begin{eqnarray}
2\sigma s\frac{d\sigma}{ds} & = & (1+\sigma)^2u^2;\\
2\frac{\sigma s}u\frac{du}{ds} & = & \sigma-1-(1+\sigma)u^2;
  \end{eqnarray}
which has the first integral
 \eqb
(1+\sigma)^2u^2-(\sigma-1)^2=c_1.
 \label{core_u}\eqe
The general solution is written as
 \begin{eqnarray}
s=c_2 \left[1-\sigma+\sqrt{-c_1}\right]^{1+1/\sqrt{-c_1}}
\left[1-\sigma-\sqrt{-c_1})\right]^{1-1/\sqrt{-c_1}};
\quad c_1<0; \label{core_sigma<1}\\
s=c_2\left[c_1+(\sigma-1)^2\right] \exp{\left[\frac
2{\sqrt{c_1}}\arctan\frac{\sigma-1}{\sqrt{c_1}}\right]}; \quad
c_1>0.
 \label{core_sigma>1}\end{eqnarray}
The solutions with $c_1<0$ describes the flow with $\sigma$
growing with the radius, and therefore with $\Psi$, until it
reaches a constant $\sigma_{0}=1-\sqrt{-c_1}<1$. Therefore
this solution represents the structure of the core in
low-$\sigma$ jets.
Transition to $\sigma\to\sigma_{0}$ is described by the
expression
 \eqb
\sigma=\sigma_0-
\left[2(1-\sigma_0)\right]^{(2-\sigma_0)/\sigma_0}\left(\frac{c_2}s\right)^{-(1-\sigma_0)/\sigma_0}.
 \eqe
Making use of Eqs. (\ref{newvar}) and (\ref{core_u}), one writes this
asymptotics in the original variables as
 \eqb
\frac{\gamma}{\gamma_{\rm in}}=\frac
1{1+\sigma_0}\frac{\Psi}{\widetilde\Psi};\quad \frac
X{\gamma_{\rm
in}}=\left[2(1-\sigma_0)\right]^{-1/\sigma_0}c_2^{(1-\sigma_0)/2\sigma_0}
\left(\frac{\Psi}{\widetilde\Psi}\right)^{(1+\sigma_0)/2\sigma_0}.
 \label{sigma<1}\eqe

The solutions with $c_1>0$ become Poynting dominated far enough
from the axis. In the limit
$\sigma\gg 1$, one finds
\eqb u=1;\qquad s=c_2\sigma^2\exp{\left(\frac
{\pi}{\sqrt{c_1}}\right)};
 \label{small_v}\eqe
or, returning to the original variables,
 \eqb
\gamma=X;\qquad
\frac{\Psi}{\widetilde{\Psi}}=c_2^{-1}\exp{\left(-\frac
{\pi}{\sqrt{c_1}}\right)}\left(\frac{X}{\gamma_{\rm in}}\right)^2.
 \label{asymp_Poynt}\eqe
One sees that far from the axis, the poloidal field becomes homogeneous
and the solution is smoothly matched with the solution for the
Poynting dominated domain, Eqs. (\ref{Phi}),
(\ref{jet_radius}) and (\ref{acceleration1}).

When $c_1$ is not small, Eq. (\ref{small_v}) implies that
$\sigma$ is large at large $s$ so that the flow is Poynting
dominated everywhere  except of the region
$\Psi\lesssim\widetilde{\Psi}$. When $c_1$ is small, the
Poynting dominated domain arises only very far from the axis,
at $s\gg\exp(\pi/\sqrt{c_1})\gg 1$. In the intermediate region,
$1\ll s\ll\exp(\pi/\sqrt{c_1})$, the solution (\ref{core_sigma>1}) is reduced to an
intermediate asymptotics
 \eqb
\sigma=1+\frac{c_1}2\ln{\frac s{c_1}}.
 \label{intermediate_sigma}\eqe
This means that in the core of the Poynting dominated jet,
$\sigma$ is close but remains larger than unity. It could
become less than unity only when the whole jet ceases to be
Poyniting dominated. In the original variables, the solution in
the intermediate zone is
 \eqb
\frac{\Psi}{\widetilde{\Psi}}=\frac{c_1^{1/2}X}{\gamma_{\rm
in}};\quad \gamma=\frac 12c_1^{1/2}X.
 \label{intermediate}\eqe


\begin{figure*}
\includegraphics[scale=0.45]{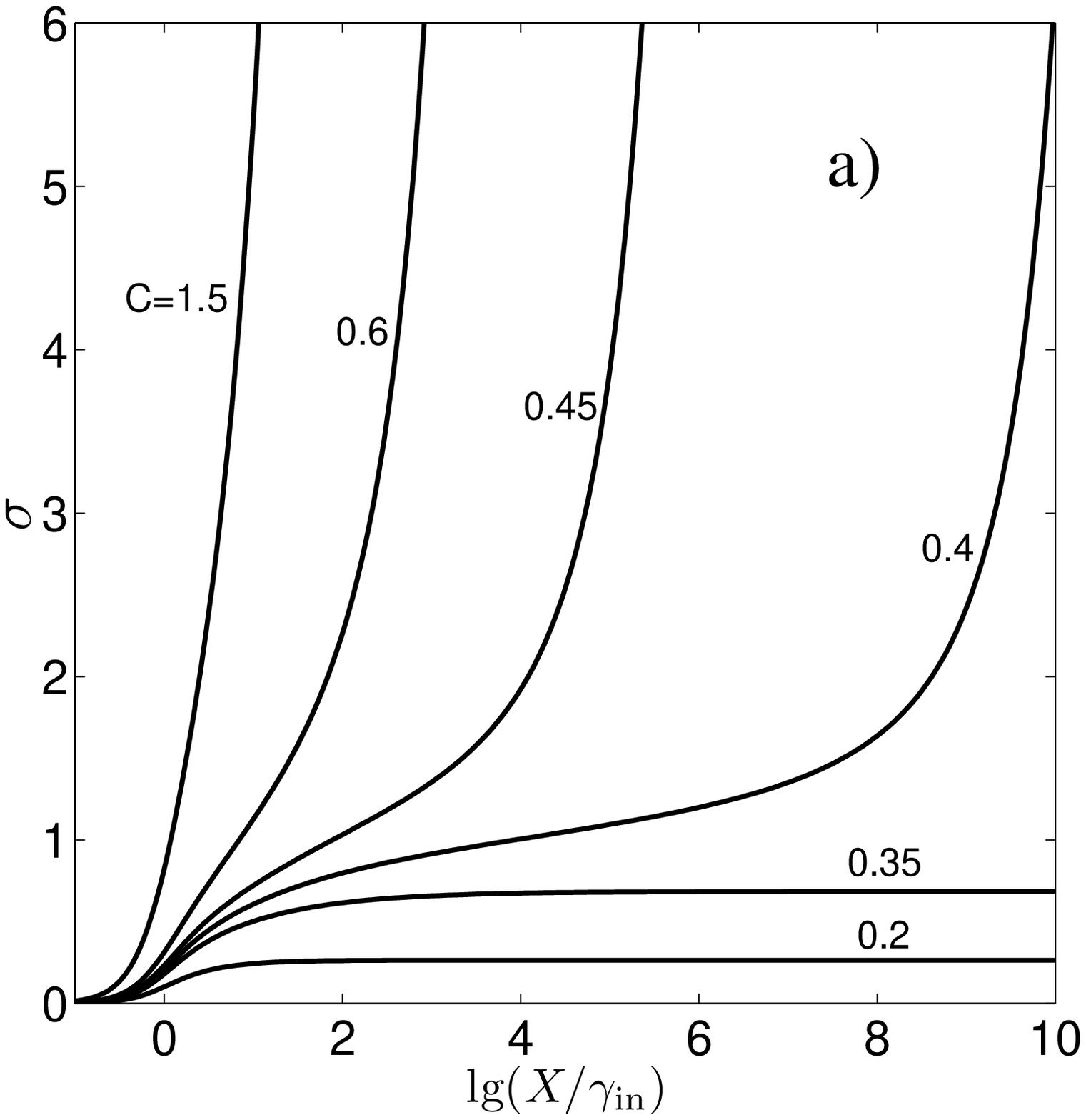}
\includegraphics[scale=0.45]{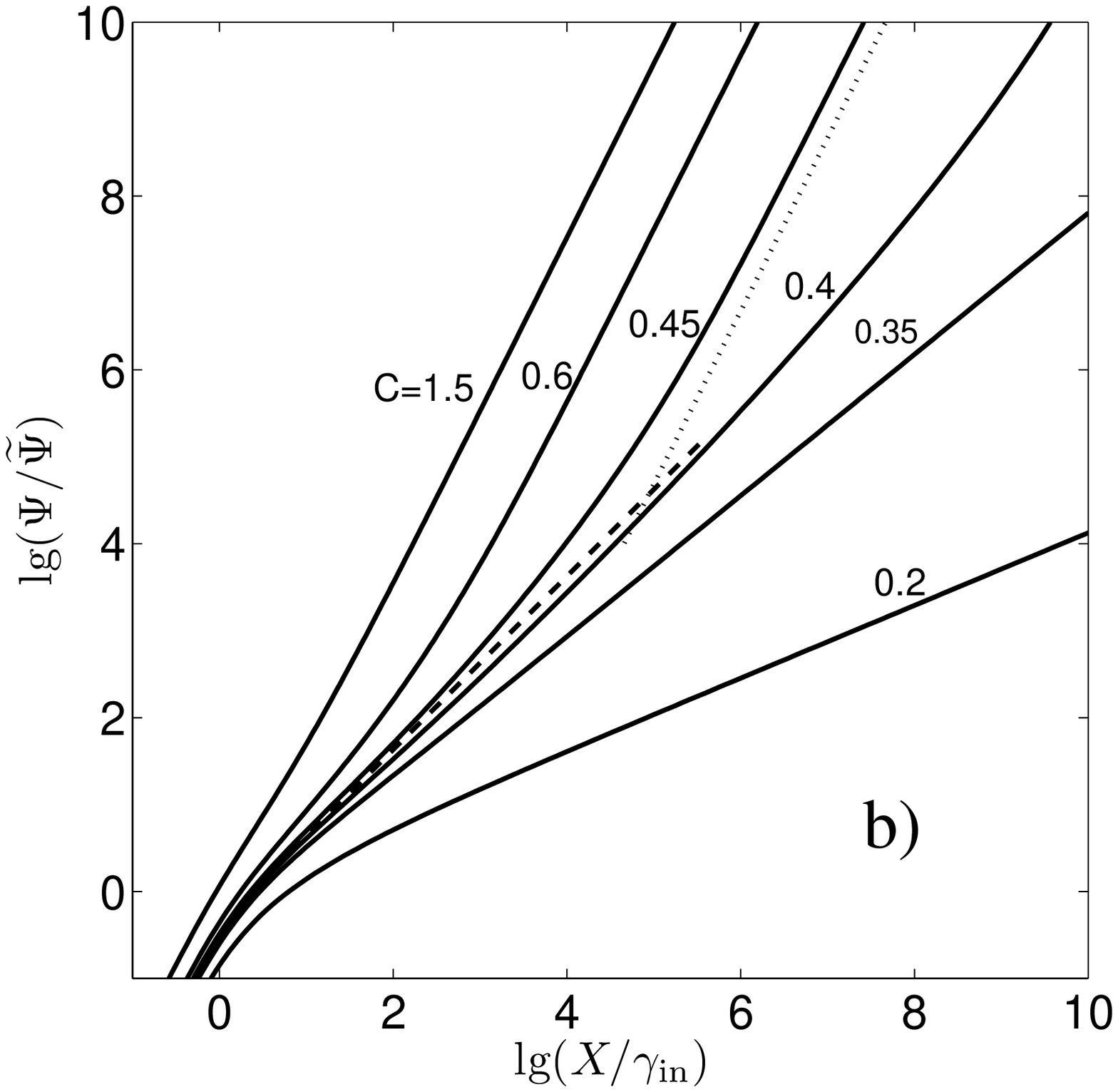}

\includegraphics[scale=0.45]{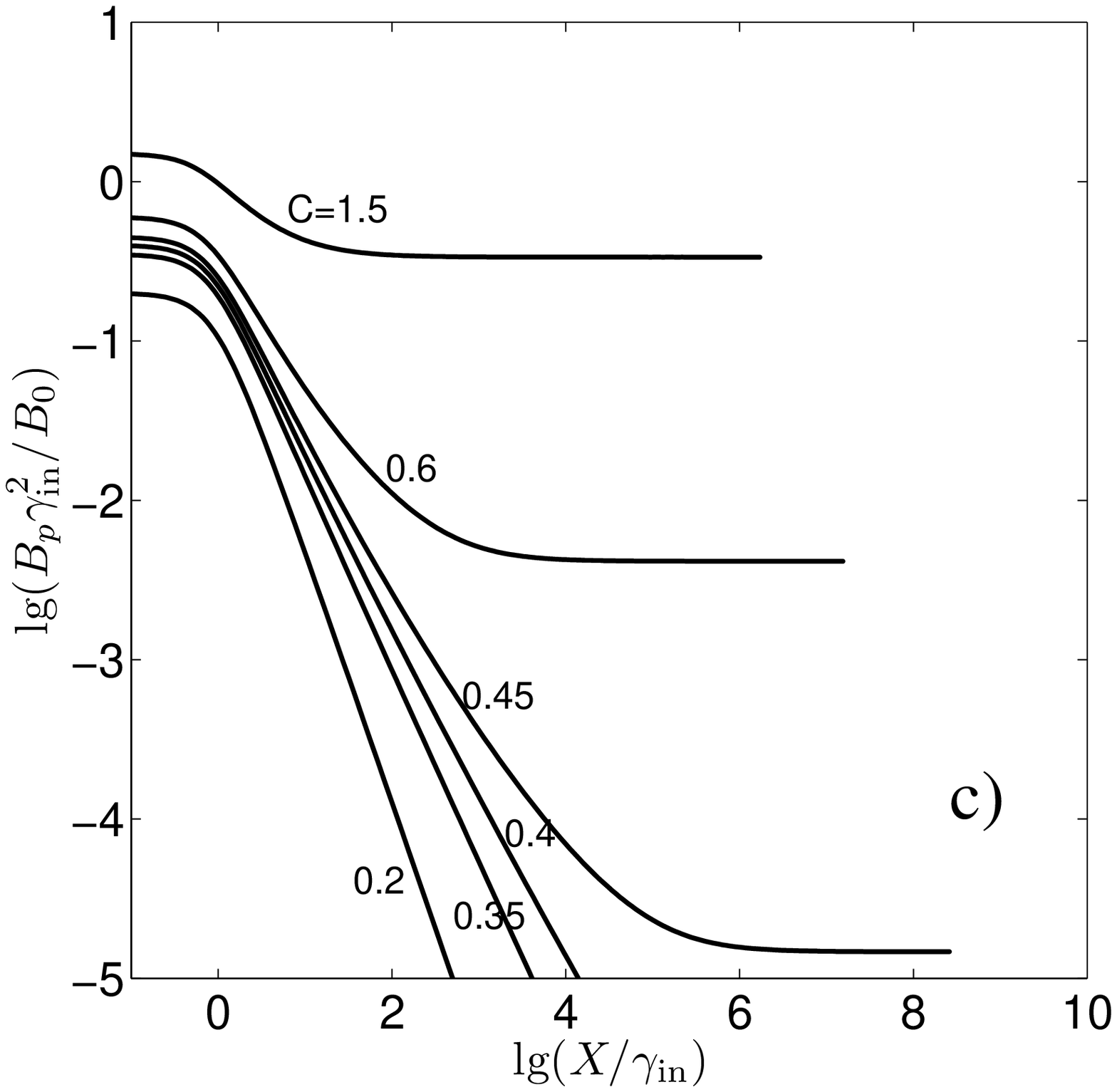}
\includegraphics[scale=0.45]{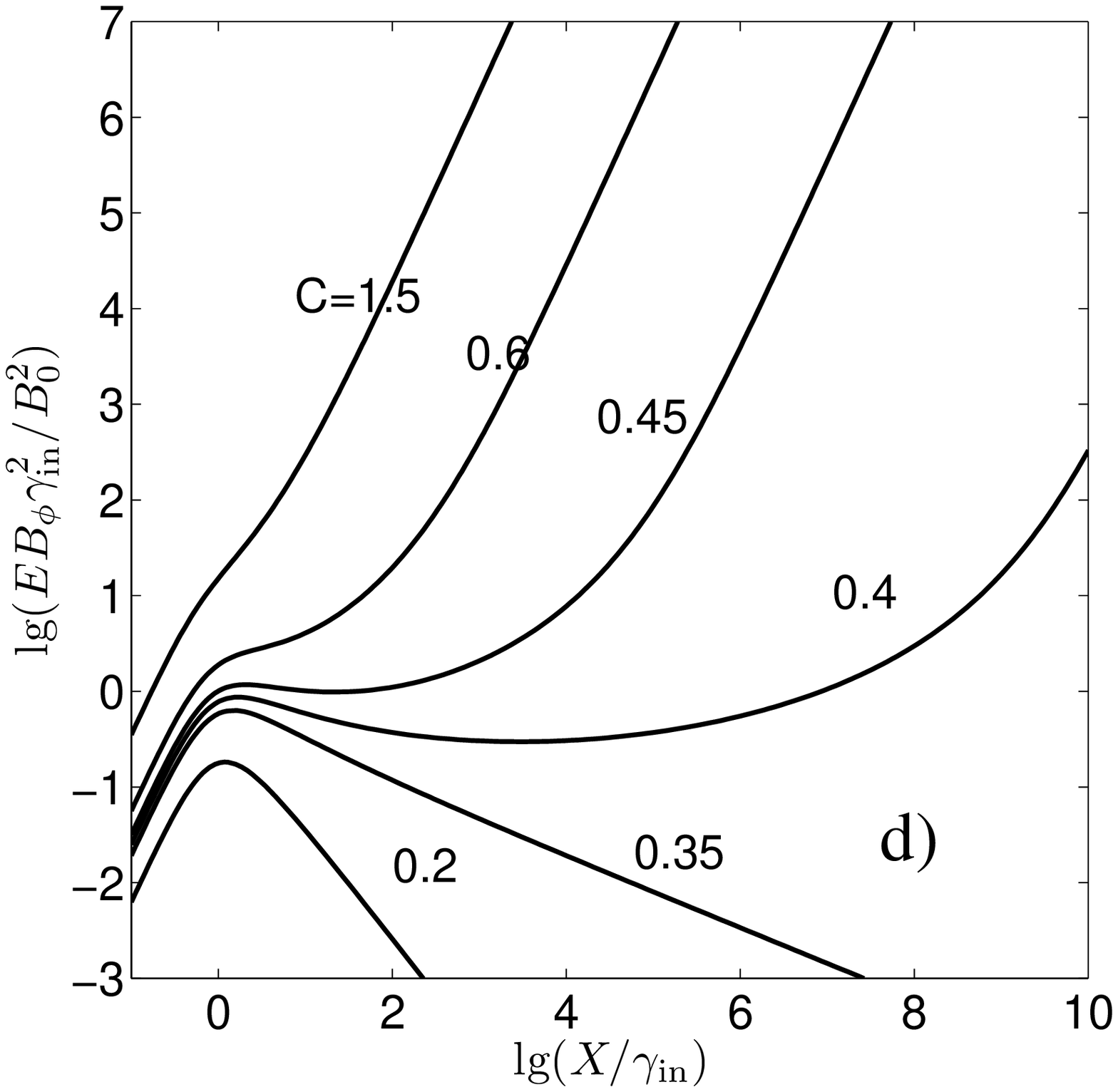}
\caption{The transverse structure of the core of the jet. Shown
are the ratio of the Poynting to the kinetic energy flux (a),
distribution of the poloidal magnetic flux (b), the poloidal
magnetic field (c) and the Poynting flux (d).  Each curve
describes the structure of the core at some distance $z$. The
curves are labeled by the constant $\cal C$ from the left
boundary condition (\ref{init}); the less $\cal C$ the larger
the corresponding $z$. The dashed and dotted lines show the
asymptotics $\Psi\propto X$ and $\Psi\propto X^2$, Eqs.
(\ref{intermediate}) and (\ref{asymp_Poynt}), correspondingly.}
\end{figure*}

Numerical solutions to Eqs. (\ref{core_tr1}) and
(\ref{core_Bern1}) are presented in Fig. 6. For all asymptotics
to be seen clearly, we plotted the curves in a very large
scale. The curves are labeled by an appropriate constant $\cal
C$ in the left boundary condition (\ref{init}). When ${\cal
C}>1.5$, the flow is Poynting dominated everywhere except of
the region $\Psi\lesssim\widetilde{\Psi}$. The poloidal
magnetic field is homogeneous as it should be in the Poynting
dominated flow with the energy integral $\mu(\Psi)$ given by
Eq. (\ref{energy1}). In the case $0.38<{\cal C}<1.5$, the
solution goes to the Poynting dominated asymptotics
(\ref{asymp_Poynt}) only at large enough distances from the
axis; $B_p$ goes to a constant in this zone. Between
$\Psi\sim\widetilde{\Psi}$ and the Poynting dominated zone, the
solution is roughly described by an intermediate asymptotics
(\ref{intermediate_sigma}) and (\ref{intermediate}). The
poloidal magnetic field varies roughly as $\propto 1/X$ in this
zone, which means that the toroidal field, $B_{\phi}=XB_p$, and
the Poynting flux remain roughly constant. At ${\cal C}<0.38$,
the solution is described, at $\xi\gg 1$, by the asymptotics
(\ref{sigma<1}), corresponding to $\sigma={\it const}<1$. In
these solutions, the poloidal magnetic field decreases faster
than $1/X$. These solutions describe the core of the jet at the stage when
most of the Poynting flux is already converted into the kinetic
energy.

Any solution to Eqs. (\ref{core_tr1}) and (\ref{core_Bern1})
with the initial condition (\ref{init}) describes the
transverse structure of the jet at some $Z$. Generally $\sigma$
decreases with $Z$ 
therefore the curves in Fig. 6
describe the $Z$ development of the jet "upside down", i.e. the
upper curves describe the transverse structure of the jet at
smaller $Z$. The moderately magnetized, $\sigma\sim 1$, zone
occupies initially only the region
$\Psi\sim\widetilde{\Psi}$. As the distance grows, the
$\sigma\sim 1$ zone extends to a larger $\Psi$.

\begin{figure*}
\includegraphics[scale=0.45]{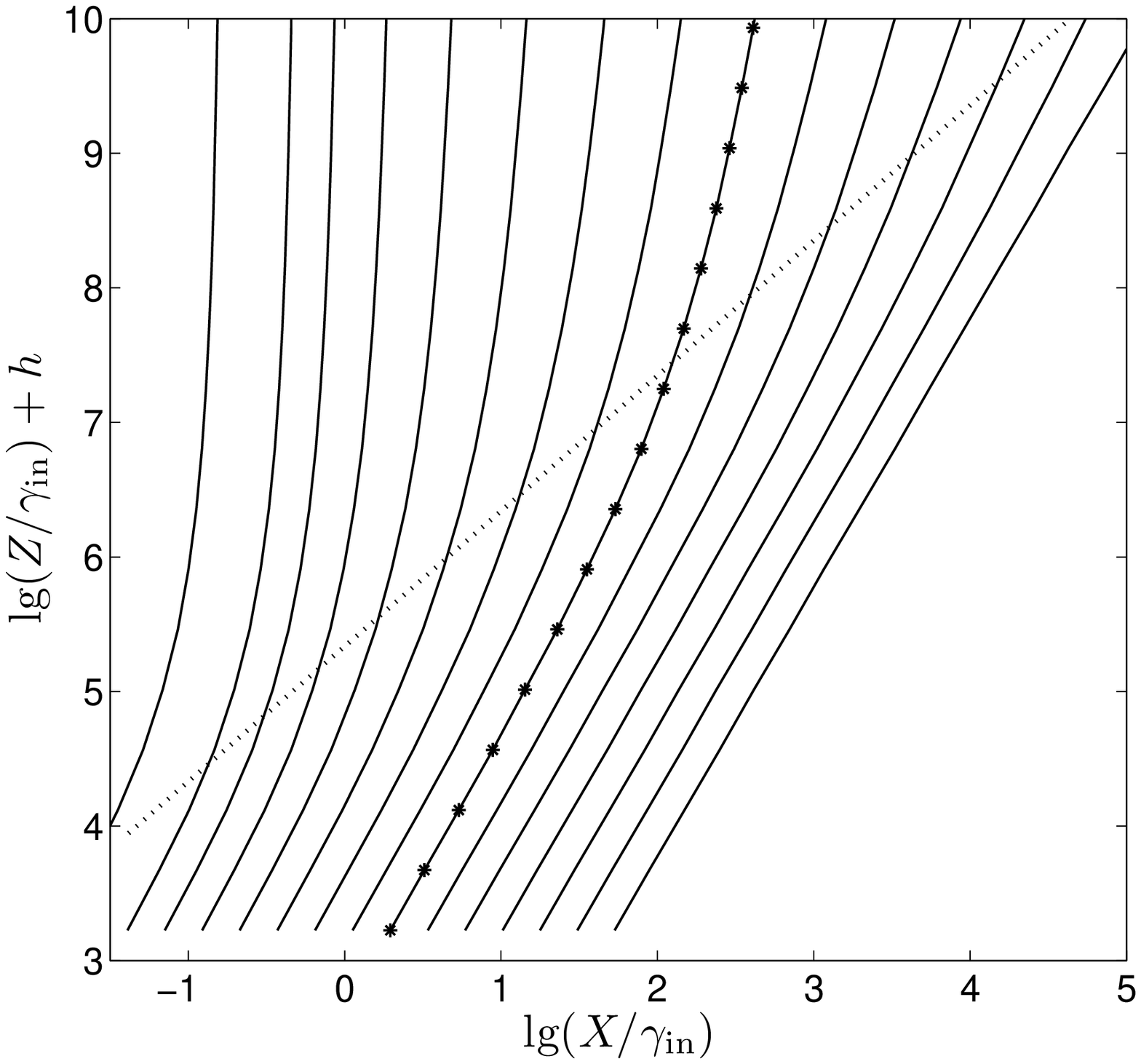}
\caption{The flux surfaces at $\kappa=2$;
$\beta>1/4$; $\gamma_{\rm max}=1.3\cdot 10^5$. The dotted line
shows the boundary of the moderately magnetized core according to Eq. (\ref{core_boundary}). $h=(1/4)\lg\left[(\beta-1/4)/\alpha\right]$}.
\end{figure*}

\begin{figure*}
\includegraphics[scale=0.45]{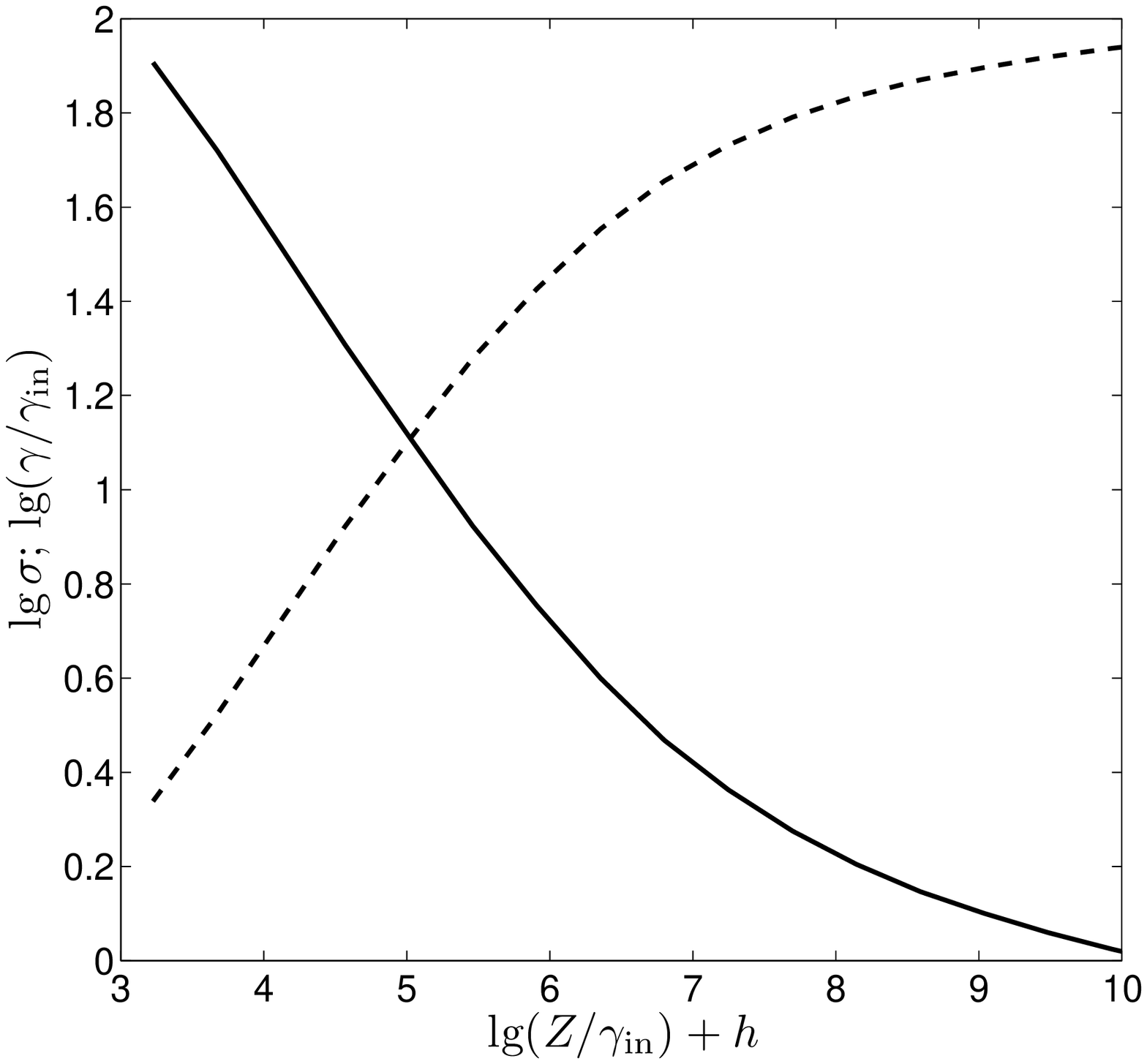}
\caption{Evolution of the Lorentz factor (dashed) and of the
ratio of the Poynting to the kinetic energy flux (solid) along
the flux surface marked by asterisks in Fig. 7.}
\end{figure*}

The structure of the core of the Poynting dominated jet may be
found in any specific case as follows. First one finds the shape
of the flux surfaces in the Poynting dominated region, $Y(Z)$,
as it was described in the previous section. Then the
transverse structure of the core at any $Z$ is described by a solution to
Eqs. (\ref{core_tr1}) and (\ref{core_Bern1}) satisfying the
left boundary condition (\ref{init}) and matched, at large
$s$, with the inner solution for the Poynting dominated flow,
Eqs. (\ref{Phi}), (\ref{jet_radius}) and (\ref{alpha_lin}). The
matching is reduced to finding an appropriate constant $C$ in
the left boundary condition (\ref{init}), which could be done
by bisection: choosing ${\cal C}_1$ and ${\cal C}_2$ such that the first solution
goes at large $\Psi$ to $r$ larger than that of Eq. (\ref{jet_radius})
whereas the second one goes to $r$ smaller than that of Eq. (\ref{jet_radius}),
one finds the solution for ${\cal C}=({\cal C}_1+{\cal C}_2)/2$ and continues until the necessary
solution is found.

As an example, we presented in Fig. 7 the
structure of the jet confined by the outer pressure ${\cal P}=\beta Z^{-2}$; $\beta>1/4$.
In the Poynting dominated region, the shape of such a jet is described by Eq. (\ref{Ykappa=2a}).
One sees that initially the flux surfaces within the jet
diverge as $Z^{1/2}$ together with the boundary of the jet but
eventually a cylindrical core is formed \citep{tomimatsu94,bogovalov95}. The transition to the
cylindrical flux surfaces occurs when the flow becomes moderately
magnetized, $\sigma\sim 1$, and the Lorentz factor saturates, see Fig. 8. The
transition is in fact very slow therefore  the flux surfaces
$\Psi\gg\widetilde{\Psi}$ become cylindrical only at extremely
large distances. The further from the axis, the later the
saturation is achieved. One sees from Fig.8 that while the main body of the jet
remains Poynting dominated, $\sigma$ in the core remains
larger than unity just approaching unity from
above, which agrees with the general analysis presented above.

\section{Conclusions}

In this paper, we developed an asymptotic theory of
relativistic, magnetized jets. The study was motivated by the
fact that acceleration and collimation of relativistic MHD
outflows occur in a very extended zone far beyond the light
cylinder. This is because the Lorentz force is nearly
compensated by the electric force when the flow speed
approaches to the speed of light. Because the dominant terms in
the the full set of MHD equations nearly cancel each other in
the far zone, it is difficult to solve them directly even
numerically. In this paper, we derived asymptotic equations,
which describe relativistic, steady state, axisymmetric MHD
flows in the far zone. These equations could be easily solved
numerically because they do not contain either intrinsic small
scales like $\Omega r$ or terms that nearly cancel each other.
Moreover, in many cases one can solve them analytically or
semi-analytically and find simple scalings, which provide
qualitative understanding of the basic properties of
relativistic MHD flows.

We applied these equations to externally confined, collimated
flows. Qualitative analysis shows that there are two regimes of
collimation, which we called equilibrium and non-equilibrium,
correspondingly. In the first regime, the flow structure at any
distance from the source is the same as the structure of an
appropriate cylindrical flow. We call this regime equilibrium
because in this case, the residual of the magnetic hoop stress
and the electrical force is balanced, as in true cylindrical
configurations, by the pressure of the poloidal magnetic field.
In the non-equilibrium regime, the pressure of the poloidal field
is negligibly small so that the flow behaves as if it possesses
purely azimuthal field. Such a flow could be conceived as
composed from coaxial magnetic loops.

The ourflow is in the equilibrium within the parabola $r^2\Omega<z$
whereas the non-equilibrium regime occurs only outside this parabola, i.e.
if the jet is not too narrow. Close enough to the axis, the flow is always
in the cylindrical equilibrium. An interesting feature is that
even though the pressure of the poloidal field does not hinder
the collimation in the non-equilibrium regime, collimation is
in fact slower in this regime than in the equilibrium one. The
reason is that one can neglect the poloidal field only if the
flow expands rapidly enough. In the two collimation regimes, the
flow is accelerated in different ways. In the equilibrium regime,
the flow Lorentz factor goes as $\gamma\sim\Omega r$ whereas in
the non-equilibrium regime, the scaling is $\gamma\sim\sqrt{{\cal
R}/r}\sim z/r$.

The shape of the flux surfaces in the Poynting dominated,
externally confined jet could be found by solving a simple
ordinary differential equation. We studied in detail the
structure of jets with a constant angular velocity confined by
the external pressure with the power law profile, $p\propto
z^{-\kappa}$. At $\kappa\le 2$, the jet acquires a parabolical
shape $r\propto z^k$, where $k<1$ depends on the pressure
profile. The jet is collimated and accelerated until the flow
ceases to be Poynting dominated. The larger the initial $\sigma$,
the larger the final Lorentz factor of the flow. The opening angle, $\Theta$,
decreases continuously so that the flow remains causally
connected, $\Theta\gamma\lesssim 1$. At
$\kappa>2$, the flow becomes asymptotically radial. If $\kappa$
only slightly exceeds 2, the flow still could be collimated
before the flow lines become straight. The final collimation
angle depends only on the pressure profile. When the flow
becomes radial, the Lorentz factor could continue to grow so that
the flow could become causally disconnected, $\Theta\gamma>1$.
However, the acceleration is practically saturated when flow
reaches the terminal Lorentz factor $\gamma_t\sim \gamma_{\rm
max}^{1/3}\Theta^{-2/3}$. This generalizes the well
known result \citep{tomimatsu94,beskin_etal98} that the non-collimated flow is
accelerated practically only to $\gamma\sim \gamma_{\rm
max}^{1/3}$.

The Poynting flux generally goes to zero at the axis of the
flow therefore a $\sigma\sim 1$ core is always presented in
the Poynting dominated jets. We have shown that as the flow is
accelerated, this core expands and the flow lines within the
core approach cylinders. At $\kappa\le 2$, the core expands
until the $\sigma\sim 1$ region eventually occupies the whole
jet whereas at $\kappa>2$, the main body of the flow remains
Poynting dominated up to logarithmically large distances.

\acknowledgements
I am grateful to Vasily Beskin and Nektarios Vlahakis for useful discussions.
This work was supported by the US-Israeli Binational Science Foundation and by the Israeli Science Foundation.

\bibliographystyle{apj}

\bibliography{mhd}

\hyphenation{Post-Script Sprin-ger}
\begin{thebibliography}{69}
\expandafter\ifx\csname natexlab\endcsname\relax\def\natexlab#1{#1}\fi

\bibitem[{{Begelman} \& {Li}(1994)}]{begelman_li94}
{Begelman}, M.~C., \& {Li}, Z.-Y. 1994, \apj, 426, 269

\bibitem[{{Beskin} {et~al.}(1998){Beskin}, {Kuznetsova}, \&
  {Rafikov}}]{beskin_etal98}
{Beskin}, V.~S., {Kuznetsova}, I.~V., \& {Rafikov}, R.~R. 1998, \mnras, 299,
  341

\bibitem[{{Beskin} \& {Malyshkin}(2000)}]{beskin_malyshkin00}
{Beskin}, V.~S., \& {Malyshkin}, L.~M. 2000, Astronomy Letters, 26, 208

\bibitem[{{Beskin} \& {Nokhrina}(2006)}]{beskin_nokhrina06}
{Beskin}, V.~S., \& {Nokhrina}, E.~E. 2006, \mnras, 367, 375

\bibitem[{{Beskin} \& {Nokhrina}(2008)}]{beskin_nokhrina08}
---. 2008, ArXiv e-prints

\bibitem[{{Beskin} {et~al.}(2004){Beskin}, {Zakamska}, \&
  {Sol}}]{beskin_etal04}
{Beskin}, V.~S., {Zakamska}, N.~L., \& {Sol}, H. 2004, \mnras, 347, 587

\bibitem[{{Bisnovatyi-Kogan} \& {Lovelace}(2007)}]{bisnovaty_lovelace07}
{Bisnovatyi-Kogan}, G.~S., \& {Lovelace}, R.~V.~E. 2007, \apjl, 667, L167

\bibitem[{{Bisnovatyi-Kogan} \& {Ruzmaikin}(1976)}]{bisnovaty_ruzmaikin76}
{Bisnovatyi-Kogan}, G.~S., \& {Ruzmaikin}, A.~A. 1976, \apss, 42, 401

\bibitem[{{Blandford}(1976)}]{blandford76}
{Blandford}, R.~D. 1976, \mnras, 176, 465

\bibitem[{{Blandford} \& {Znajek}(1977)}]{blandford_znajek77}
{Blandford}, R.~D., \& {Znajek}, R.~L. 1977, \mnras, 179, 433

\bibitem[{{Bogovalov} \& {Tsinganos}(1999)}]{bogovalov_tsinganos99}
{Bogovalov}, S., \& {Tsinganos}, K. 1999, \mnras, 305, 211

\bibitem[{{Bogovalov}(1995)}]{bogovalov95}
{Bogovalov}, S.~V. 1995, Astronomy Letters, 21, 565

\bibitem[{{Bogovalov}(1997)}]{bogovalov97}
---. 1997, \aap, 323, 634

\bibitem[{{Bogovalov}(1998)}]{bogovalov98}
---. 1998, Astronomy Letters, 24, 321

\bibitem[{{Bromberg} \& {Levinson}(2007)}]{bromberg_levinson07}
{Bromberg}, O., \& {Levinson}, A. 2007, \apj, 671, 678

\bibitem[{{Buckley}(1977)}]{buckley77}
{Buckley}, R. 1977, \mnras, 180, 125

\bibitem[{{Camenzind}(1986)}]{camenzind86}
{Camenzind}, M. 1986, \aap, 162, 32

\bibitem[{{Chiueh} {et~al.}(1991){Chiueh}, {Li}, \&
  {Begelman}}]{chiueh_li_begelman91}
{Chiueh}, T., {Li}, Z.-Y., \& {Begelman}, M.~C. 1991, \apj, 377, 462

\bibitem[{{Chiueh} {et~al.}(1998){Chiueh}, {Li}, \& {Begelman}}]{chieh_etal98}
---. 1998, \apj, 505, 835

\bibitem[{{Contopoulos} {et~al.}(1999){Contopoulos}, {Kazanas}, \&
  {Fendt}}]{contopoulos_etal99}
{Contopoulos}, I., {Kazanas}, D., \& {Fendt}, C. 1999, \apj, 511, 351

\bibitem[{{Contopoulos}(1995)}]{contopoulos95}
{Contopoulos}, J. 1995, \apj, 446, 67

\bibitem[{{Drenkhahn}(2002)}]{drenkhahn02}
{Drenkhahn}, G. 2002, \aap, 387, 714

\bibitem[{{Drenkhahn} \& {Spruit}(2002)}]{drenkhahn_spruit02}
{Drenkhahn}, G., \& {Spruit}, H.~C. 2002, \aap, 391, 1141

\bibitem[{{Eichler}(1982)}]{eichler82}
{Eichler}, D. 1982, \apj, 263, 571

\bibitem[{{Eichler}(1993)}]{eichler93}
---. 1993, \apj, 419, 111

\bibitem[{{Fendt}(1997)}]{fendt97}
{Fendt}, C. 1997, \aap, 319, 1025

\bibitem[{{Gourgouliatos} \& {Lynden-Bell}(2008)}]{gourgouliatos_lynden-bell08}
{Gourgouliatos}, K.~N., \& {Lynden-Bell}, D. 2008, \mnras, 391, 268

\bibitem[{{Heyvaerts} \& {Norman}(1989)}]{heyvaerts_norman89}
{Heyvaerts}, J., \& {Norman}, C. 1989, \apj, 347, 1055

\bibitem[{{Heyvaerts} \& {Norman}(2003)}]{heyvaerts_norman03}
---. 2003, \apj, 596, 1240

\bibitem[{{Kato} {et~al.}(2004){Kato}, {Mineshige}, \& {Shibata}}]{kato_etal04}
{Kato}, Y., {Mineshige}, S., \& {Shibata}, K. 2004, \apj, 605, 307

\bibitem[{{Kirk} {et~al.}(2007){Kirk}, {Lyubarsky}, \& {Petri}}]{kirk_etal07}
{Kirk}, J.~G., {Lyubarsky}, Y., \& {Petri}, J. 2007, ArXiv Astrophysics
  e-prints

\bibitem[{{Kirk} \& {Skj{\ae}raasen}(2003)}]{kirk_olaf03}
{Kirk}, J.~G., \& {Skj{\ae}raasen}, O. 2003, \apj, 591, 366

\bibitem[{{Komissarov} {et~al.}(2008){Komissarov}, {Vlahakis}, {Konigl}, \&
  {Barkov}}]{komissarov_etal08}
{Komissarov}, S., {Vlahakis}, N., {Konigl}, A., \& {Barkov}, M. 2008, ArXiv
  e-prints

\bibitem[{{Komissarov} {et~al.}(2007){Komissarov}, {Barkov}, {Vlahakis}, \&
  {K{\"o}nigl}}]{komissarov_etal07}
{Komissarov}, S.~S., {Barkov}, M.~V., {Vlahakis}, N., \& {K{\"o}nigl}, A. 2007,
  \mnras, 380, 51

\bibitem[{{Levinson} \& {Eichler}(2000)}]{levinson_eichler00}
{Levinson}, A., \& {Eichler}, D. 2000, Physical Review Letters, 85, 236

\bibitem[{{Li} {et~al.}(1992){Li}, {Chiueh}, \&
  {Begelman}}]{li_chiueh_begelman92}
{Li}, Z.-Y., {Chiueh}, T., \& {Begelman}, M.~C. 1992, \apj, 394, 459

\bibitem[{{Lovelace}(1976)}]{lovelace76}
{Lovelace}, R.~V.~E. 1976, \nat, 262, 649

\bibitem[{{Lovelace} {et~al.}(2002){Lovelace}, {Li}, {Koldoba}, {Ustyugova}, \&
  {Romanova}}]{lovelace_etal02}
{Lovelace}, R.~V.~E., {Li}, H., {Koldoba}, A.~V., {Ustyugova}, G.~V., \&
  {Romanova}, M.~M. 2002, \apj, 572, 445

\bibitem[{{Lovelace} {et~al.}(1986){Lovelace}, {Mehanian}, {Mobarry}, \&
  {Sulkanen}}]{lovelace_etal86}
{Lovelace}, R.~V.~E., {Mehanian}, C., {Mobarry}, C.~M., \& {Sulkanen}, M.~E.
  1986, \apjs, 62, 1

\bibitem[{{Lovelace} \& {Romanova}(2003)}]{lovelace_romanova03}
{Lovelace}, R.~V.~E., \& {Romanova}, M.~M. 2003, \apjl, 596, L159

\bibitem[{{Lovelace} {et~al.}(2006){Lovelace}, {Turner}, \&
  {Romanova}}]{lovelace_etal06}
{Lovelace}, R.~V.~E., {Turner}, L., \& {Romanova}, M.~M. 2006, \apj, 652, 1494

\bibitem[{{Lynden-Bell}(1996)}]{lynden-bell96}
{Lynden-Bell}, D. 1996, \mnras, 279, 389

\bibitem[{{Lynden-Bell}(2006)}]{lynden-bell06}
---. 2006, \mnras, 369, 1167

\bibitem[{{Lyubarsky} \& {Eichler}(2001)}]{lyubarsky_eichler01}
{Lyubarsky}, Y., \& {Eichler}, D. 2001, \apj, 562, 494

\bibitem[{{Lyubarsky} \& {Kirk}(2001)}]{lyubarsky_kirk01}
{Lyubarsky}, Y., \& {Kirk}, J.~G. 2001, \apj, 547, 437

\bibitem[{{Michel}(1969)}]{michel69}
{Michel}, F.~C. 1969, \apj, 158, 727

\bibitem[{{Nakamura} {et~al.}(2006){Nakamura}, {Li}, \& {Li}}]{nakamura_etal06}
{Nakamura}, M., {Li}, H., \& {Li}, S. 2006, \apj, 652, 1059

\bibitem[{{Nakamura} {et~al.}(2007){Nakamura}, {Li}, \& {Li}}]{nakamura_etal07}
---. 2007, \apj, 656, 721

\bibitem[{{Narayan} {et~al.}(2007){Narayan}, {McKinney}, \&
  {Farmer}}]{narayan_etal07}
{Narayan}, R., {McKinney}, J.~C., \& {Farmer}, A.~J. 2007, \mnras, 375, 548

\bibitem[{{Okamoto}(1974)}]{okamoto74}
{Okamoto}, I. 1974, \mnras, 167, 457

\bibitem[{{Okamoto}(1978)}]{okamoto78}
---. 1978, \mnras, 185, 69

\bibitem[{{Peter} \& {Eichler}(1995)}]{peter_eichler95}
{Peter}, W., \& {Eichler}, D. 1995, \apj, 438, 244

\bibitem[{{Polyanin} \& {Zaitsev}(2002)}]{polyanin_zaitsev02}
{Polyanin}, A.~D., \& {Zaitsev}, V.~F. 2002, {Handbook of Exact Solutions for
  Ordinary Differential Equations} (Chapman \& Hall/CRC, 2002, 787 p.)

\bibitem[{{Rothstein} \& {Lovelace}(2008)}]{rothstein_lovelace08}
{Rothstein}, D.~M., \& {Lovelace}, R.~V.~E. 2008, \apj, 677, 1221

\bibitem[{{Sherwin} \& {Lynden-Bell}(2007)}]{lynden-bell07}
{Sherwin}, B.~D., \& {Lynden-Bell}, D. 2007, \mnras, 378, 409

\bibitem[{{Spruit} {et~al.}(1997){Spruit}, {Foglizzo}, \&
  {Stehle}}]{spruit_etal97}
{Spruit}, H.~C., {Foglizzo}, T., \& {Stehle}, R. 1997, \mnras, 288, 333

\bibitem[{{Tchekhovskoy} {et~al.}(2008){Tchekhovskoy}, {McKinney}, \&
  {Narayan}}]{tchekhovskoy08}
{Tchekhovskoy}, A., {McKinney}, J.~C., \& {Narayan}, R. 2008, \mnras, 388, 551

\bibitem[{{Tchekhovskoy} {et~al.}(2009){Tchekhovskoy}, {McKinney}, \&
  {Narayan}}]{tchekhovskoy09}
---. 2009, ArXiv e-prints

\bibitem[{{Thompson}(1994)}]{thompson94}
{Thompson}, C. 1994, \mnras, 270, 480

\bibitem[{{Timokhin}(2006)}]{timokhin06}
{Timokhin}, A.~N. 2006, \mnras, 368, 1055

\bibitem[{{Tomimatsu}(1994)}]{tomimatsu94}
{Tomimatsu}, A. 1994, \pasj, 46, 123

\bibitem[{{Tsinganos} {et~al.}(1996){Tsinganos}, {Sauty}, {Surlantzis},
  {Trussoni}, \& {Contopoulos}}]{tsinganos_etal96}
{Tsinganos}, K., {Sauty}, C., {Surlantzis}, G., {Trussoni}, E., \&
  {Contopoulos}, J. 1996, \mnras, 283, 811

\bibitem[{{Uzdensky}(2004)}]{uzdensky04}
{Uzdensky}, D.~A. 2004, \apj, 603, 652

\bibitem[{{Uzdensky}(2005)}]{uzdensky05}
---. 2005, \apj, 620, 889

\bibitem[{{Uzdensky} \& {MacFadyen}(2006)}]{uzdensky_macfadyen06}
{Uzdensky}, D.~A., \& {MacFadyen}, A.~I. 2006, \apj, 647, 1192

\bibitem[{{Vlahakis}(2004)}]{vlahakis04}
{Vlahakis}, N. 2004, \apj, 600, 324

\bibitem[{{Vlahakis} \& {K{\"o}nigl}(2003{\natexlab{a}})}]{vlahakis_konigl03a}
{Vlahakis}, N., \& {K{\"o}nigl}, A. 2003{\natexlab{a}}, \apj, 596, 1080

\bibitem[{{Vlahakis} \& {K{\"o}nigl}(2003{\natexlab{b}})}]{vlahakis_konigl03b}
---. 2003{\natexlab{b}}, \apj, 596, 1104

\bibitem[{{Vlahakis} {et~al.}(2000){Vlahakis}, {Tsinganos}, {Sauty}, \&
  {Trussoni}}]{vlahakis_etal00}
{Vlahakis}, N., {Tsinganos}, K., {Sauty}, C., \& {Trussoni}, E. 2000, \mnras,
  318, 417

\end{thebibliography}

\end{document}